%Paper: nucl-th/9304002
%From: DEPACE@to.infn.it (Arturo De Pace - INFN Torino - +39(11)6527243)
%Date: Tue, 6 Apr 1993 15:15:00 +0200 (WET-DST)

% plain TeX
\def\reff#1{${}^{#1}$)}
\def\onee{1\hskip-.27em {\rm l}}
\def\mus{\mu_\sst{S}}
\def\muv{\mu_\sst{V}}
\def\sst#1{{\scriptscriptstyle #1}}
\magnification=\magstep1

\font\title=cmbx10 scaled\magstep1
\font\myit=cmti7 scaled\magstep1
\font\myrm=cmr9
\hsize=5in \hoffset=1cm
\hfuzz=50pt
\baselineskip=18pt
\headline{\hfil -- \folio -- \hfil}
\footline{\hfill}
\def\fakebold#1{\leavevmode\setbox0=\hbox{#1}%
  \kern-.025em\copy0 \kern-\wd0
  \kern .05em\copy0 \kern-\wd0
  \kern-.025em\raise.0433em\box0}
\def\ref#1{[#1]}

\def\bfg#1{\fakebold{$#1$}}

\def\qb{\bfg{q}}

\def\qbh{\bfg{\hat q}}
\def\kbh{\bfg{\hat k}}
\def\qh{\hat q}
\def\kh{\hat k}
\def\pb{\bfg{p}}

\def\rb{\bfg{r}}

\def\kb{\bfg{k}}

\def\taub{\bfg{\tau}}
\def\sgb{\bfg{\sigma}}
\def\sigb{\bfg{\sigma}}

\def\dag{\dagger}

\def\bt{\beta}
\def\btb{\bfg{\beta}}
\def\gm{\gamma}

\def\dt{\delta}

\def\om{\omega}

\def\limit#1{\mathrel{\rlap{\hbox{$\longrightarrow$}}
  \lower8pt\hbox{$\scriptstyle #1$}}}
\def\FFp{\Gamma_{\pi}}
\def\lm{\lambda}
\def\Lm{\Lambda}
\def\veps{\varepsilon}
\def\ums{{\scriptstyle1\over\scriptstyle2}}
\def\xia{{\xi_A}}
\def\cou{{f_\pi^2\over m_\pi^2}}
\def\kf{k_F}
\def\kv{\kb}
\def\qv{\qb}
\def\pv{\pb}
\def\h2{{ }}

\def\kin{{\h2 |Q^2|\over{2\M}}-{\h2\qv\cdot\kv\over \M}}
\def\M{m_{\scriptstyle N}}
\def\muq{{m_{\pi}^2}}
\def\hla{{\hat\lambda}}
\def\mpi{\mu_{\pi}}
\def\z{\beta}
\def\lam{\lambda_{\pi}}
\def\betv{\btb}
\def\rv{{\bfg{\kappa}}}
\def\kina{{\left|\kappa^2-\lambda^2\right|-\rv\cdot\betv}}
\def\dl{\delta}
\def\th{\theta}

\def\rapp{\mathrel{\mathop{>}\limits_{{}^\sim}}}
\def\Jb{{\bf J}}
\def\Fc{{\cal F}}
\def\Gc{{\cal G}}
\def\Hc{{\cal H}}

\def\Mc{{\cal M}}
\def\Nc{{\cal N}}
\def\Tc{{\cal T}}
\def\psibar{\bar{\psi}_r}
\def\lmt{\widetilde{\lambda}_\pi}
\def\mpt{\widetilde{m}_\pi}

\pageno=0
\centerline{\title The pion in electromagnetic and weak neutral current}
\centerline{\title nuclear response functions\footnote{*}{\rm This work is
supported in part by funds provided by the U.S. Department of Energy
(D.O.E.) under contract \#DE-AC02-76ER03069.}}
\bigskip \medskip
\centerline{W.M. Alberico$^{(\dag)}$, M.B. Barbaro$^{(\dag)}$,
A. De Pace$^{(\dag)}$}
\centerline{T.W. Donnelly$^{(\ddag)}$ and A. Molinari$^{(\dag,\S)}$}

\bigskip \medskip
\centerline{\myit $^{(\dag)}$Dipartimento di Fisica Teorica}
\centerline{\myit dell'Universit\`a di Torino}
\centerline{\myit and}
\centerline{\myit INFN, Sezione di Torino}
\centerline{\myit via P.Giuria 1, 10125 Torino, Italy}
\smallskip
\centerline{\myit $^{(\ddag)}$ Center for Theoretical Physics,}
\centerline{\myit Laboratory for Nuclear Science and Department of Physics}
\centerline{\myit Massachusetts Institute of Technology, }
\centerline{\myit Cambridge, MA 02139, USA}
\medskip
\centerline{\myit $^{(\S)}$ Ministero degli Affari Esteri--Consolato Generale
d'Italia}
\centerline{\myit Boston, MA 02116, USA}
\vskip 1.5in
\centerline{Submitted to: {\it Nuclear Physics A}}
\vfill
\vskip -12pt
\noindent CTP\#2194 \hfill March 1993
\eject
\centerline{\bf Abstract}
\smallskip
\noindent
The impact of pionic correlations and meson--exchange currents in
determining the (vector) response functions for electroweak
quasielastic lepton scattering from nuclei is discussed.  The
approach taken builds on previous work where the Fermi gas model is
used to maintain consistency in treating forces and currents (gauge
invariance) and to provide a Lorentz covariant framework.  Results
obtained in first--order perturbation theory are compared with
infinite--order summation schemes (HF and RPA) and found to
provide quite successful approximations for the quasielastic response
functions.  The role of pionic correlations in hardening the
responses $R_L$ and $R_T$ is investigated in some detail, including
studies of the relative importance of central and tensor pieces of
the force and of exchange and self--energy diagrams; in
addition, their role in significantly modifying the
longitudinal parity--violating response $R_{AV}^L$ is explored. The
MEC are shown to provide a small, but non--negligible, contribution
in determining the vector responses.
\vfill\eject

\leftline{\bf 1. Introduction}
\smallskip
The role of the pion in nuclear structure has been explored
for some time now~\reff{1} and
clear signatures of its presence as a carrier of the force
acting between nucleons via the one--pion--exchange potential (OPEP)
have been found, for example, in studies of NN--scattering~\reff{2}
or of the deuteron (see ref.~\reff{3} for a review in this area)
and as a carrier of an
electromagnetic current in np radiative capture~\reff{4} or in
the electrodisintegration of the
deuteron~\reff{5,6}. The properties of the 3--body nuclei $^3$He and $^3$H
also appear to be represented quite well by hadronic descriptions which
include pionic degrees of freedom in forces and currents, except perhaps
at the highest momentum transfers in electron scattering where extremely
short distance scales are being probed.  Importantly, this success requires
that the theory take into account 2--body meson--exchange currents (MEC) and
3--body forces (see, {\it e.g.,\/} refs.~\reff{7,8}), both having parts that
arise from pion exchange.
Such observations regarding 2-- and 3--body nuclei support
the description of nuclei in general within the framework of a hadronic
quantum field theory (the nucleus viewed as a composite system of baryons
--- especially nucleons --- and mesons --- especially pions).

However, when one comes to consider heavier nuclei the situation then
becomes less clear because of the complexity of the nuclear
many--body problem and the need to deal with less developed nuclear
wave functions than are available for the few--body nuclei.
To mention just a few examples where pionic effects have been sought,
on the one hand the enhancement of the
dipole photoabsorption sum rule over the canonical Thomas--Reiche--Kuhn
value has been interpreted with some success in
terms of a pionic contribution stemming mostly from the tensor force in
second order~\reff{9}.
On the other hand, evidence for
 the long--sought--after softening of the quasielastic peak
in the spin--longitudinal $\sigb\cdot\qb$ isovector nuclear response, a
precursor of pion condensation at nuclear density, has never been
convincingly found~\reff{10,11}.  Furthermore, while
MEC effects in electron
scattering from A$>3$ nuclei are expected (see, {\it e.g.,\/} ref.~\reff{12}),
they have been difficult to
isolate from uncertainties in nuclear structure.

In contemporary descriptions of many--body nuclear structure using
hadronic degrees of freedom different starting points have been assumed.
For example, in the relativistic mean--field approach of
Walecka and collaborators~\reff{13} (so--called relativistic quantum
hadrodynamics, QHD) in lowest--order
the pionic degrees of freedom do not enter; in that case, symmetry
considerations (translational invariance) prevent the
pion from contributing to the energy of the system.
On the other hand, variational approaches (see refs.~\reff{7,8}) suggest
an important role for the pion: for example, in the work reported in
ref.~\reff{14} it is found that over half of the
nuclear matter binding energy may be ascribed to the pion.

As the above issues are not yet fully resolved, we are left with some
uncertainty about the size of the role played by pionic degrees of
freedom in many--body nuclear structure.  When attempting to elucidate this
issue, it should be kept in mind that these degrees of
freedom may be of rather different importance in describing different
nuclear observables.  In particular, their role in accounting for the
binding energy of nuclei may be rather
different from that involved in describing
the electroweak quasielastic nuclear response functions which are the
main focus of the present work.  In the former, (particle--particle)
interactions between
nucleons in the Fermi sea are presumed to provide most of the binding
energy, whereas in the latter particle--hole excitations involving
particles typically at several hundred MeV in the continuum are
dominant.

In this paper, which enlarges on a previous one~\reff{15},
we do not aim at establishing  a relativistic quantum field theory
of nuclei based on hadronic degrees of freedom or to explore
ground--state properties using sophisticated
nuclear wave functions. Rather we treat pionic degrees of freedom as
perturbations, building on a simple, tractable, covariant nuclear model for
quasielastic response functions, namely, the relativistic Fermi
gas (RFG) model~\reff{16}.  The free RFG model appears to describe nuclear
excitations at high three--momentum transfer $q$ and high energy--transfer
$\omega$ reasonably well for kinematics near the quasielastic peak (QEP),
that is, where $\omega\approx |Q^2|/2m_N$, with $Q^2=\omega^2-q^2$.  To
the extent that pionic correlations and pionic MEC
provide only moderate corrections to the free RFG responses, one
may hope that a perturbative treatment is meaningful.  Accordingly, we
start with all Feynman diagrams carrying one pionic line and later extend
our analysis to account for perturbative diagrams up to infinite
order to test their importance. Issues relating
to Lorentz covariance and electroweak gauge invariance
can be explored in this model where a high level of consistency in
treating forces and currents can be maintained.  Some of the issues relating
to gauge invariance in this pionic model are addressed in the present work.
In subsequent work it will be
important to explore the roles played by mesons heavier than the pion
within the same framework,  however, the present scope has been
restricted to a study of the pionic effects in quasielastic electroweak
nuclear response functions.  Given the long--ranged character of pion
exchange we believe that such effects provide a natural starting point
for a more ambitious study.

In the present work we study the pionic correlation effects embodied in
the so--called self--energy and exchange diagrams, whereas,
for the MEC contributions we shall mostly rely on past work~\reff{15},
paying special attention to the problem of
fulfilling the continuity equation in achieving a
consistent treatment of currents and forces.
In accord with those previous studies, we find sizable pionic
contributions
to the electromagnetic longitudinal (spin scalar, $\sigma=0$) and transverse
(spin vector, $\sigma=1$) nuclear responses. In both cases it is found that
the correlation effects produce a hardening of the responses, that is,
a shift of the strength to higher $\omega$.

In particular, as discussed in detail in the following sections, from our
past and present analyses the following points emerge:
\smallskip
\item{a)} In isospace the contribution of the self--energy diagram to
the charge response is almost equally
split between isoscalar ($\tau=0$) and
isovector ($\tau=1$) components. (The latter, on the other hand,
is of course overwhelming in the transverse response, due to the
dominance of the isovector magnetic moment.)
In contrast, the $\tau=0$ part of
the exchange diagram is three times as large as the $\tau=1$ one
in the charge response and this
imbalance, which becomes even stronger in higher orders of perturbation
theory, is further strengthened by the difference between the isoscalar and
isovector form factors (see later).
The isoscalar dominance of the pionic exchange correlations
has dramatic consequences for the weak neutral current longitudinal
response function, as we shall see in sect.~4.
\smallskip
\item{b)} While the tensor component of the OPEP never
contributes to the self--energy in a translationally invariant system,
the exchange diagram gets a tensor
contribution, {\it but only in the transverse channel} and mostly
via the backward--going graphs.
This implies a different role for
the pionic force in the two electromagnetic responses, a finding
which should be tested against experiment.
\smallskip
\item{c)} When we extend our analysis from first to
infinite order of perturbation theory, thus generating the Hartree--Fock
(HF) approximation with the self--energy diagram
and the random phase approximation
(RPA) with the exchange diagram, our results do not change substantially.
It therefore appears that, although the pionic interaction is strong,
nevertheless for quasielastic kinematics
its effects are reasonably small at high--$q$, thus rendering perturbation
theory quite accurate already at the lowest order, at least for the
classes of diagrams studied here.
\smallskip
\item{d)} Finally, the contribution of the central component of the
pionic interaction to the exchange diagram stems largely from the
$\delta$--force and not from the finite--range one.
In keeping with the usual approach taken in studies of pionic effects, we
include a $\pi$NN vertex function $\Gamma_{\pi}$,
whose scale is set by a mass parameter $\Lambda_\pi$, to smear out the
$\delta$--piece of OPEP.  Consequently, one of the goals in the present
work is to explore the sensitivity of the resulting nuclear
quasielastic responses to the choice of the phenomenological
parameter $\Lambda_\pi$.
The question then arises: To what extent is the continuity equation
modified by the presence of $\Gamma_{\pi}$? We address this issue later
in the present work and here simply recall
that for pointlike nucleons the continuity
 equation is indeed obeyed by the OPEP
and by the related pionic MEC~\reff{12}, but is not
affected by the
$\delta$--interaction. This conclusion must be modified,
however, when the vertex function
$\Gamma_{\pi}$ is introduced, although, as discussed below,
additional MEC can be introduced~\reff{17} in a way such that the
continuity equation is still satisfied.
\smallskip
\item{} In contrast to the exchange diagram, the self--energy term gets a
contribution only from momentum--dependent forces, and therefore an unmodified
pionic $\delta$--interaction does not contribute in this channel;
in fact, the contribution coming from the $\delta$--function in OPEP, when
modified by the $\pi$NN
vertex form factor, contains an effective
momentum--dependence and so is nonzero.

Regarding the two parameters that characterize
our approach, namely the Fermi momentum $k_F$ and
the mass parameter $\Lambda_\pi$, we note that
the value of the first is essentially set by the nuclear density.
Ultimately, given extensive evaluation of all nuclear model dependences
(correlation effects, MEC effects, {\it etc.}), electron scattering
measurements in the quasielastic region should serve to determine $k_F$.
Likewise, some information on the range of acceptable values for the
phenomenological parameter $\Lambda_\pi$ can in principle be obtained from
comparison with experiment.  Clearly the value deduced for
this parameter is model--dependent: it is meant to absorb effects from
hadronic physics other than the explicit pionic degrees of freedom (to the
extent that this is even possible through a single vertex function
$\Gamma_\pi$).  In an extended model with other active mesonic degrees of
freedom presumably the physics embodied in this way can be quite different.
Even given the limited scope of the present pionic model, it is
interesting to
explore the sensitivity in our results to the
actual choice of $\Lambda_\pi$ and we do so in sect.~5.
The unexpected (and perhaps important) finding in this connection is
the following: the diagrams which contribute most to the nuclear responses are
those that are least affected by $\Gamma_{\pi}$.

With respect to the relativistic aspect of the present approach,
one limitation of our treatment arises from the static character of
our pionic interaction, while another arises from
the not--fully--relativistic character of the
fermion propagators that we use.
However we have been able to achieve  an almost exact relativistic treatment
of the fermion kinematics. As discussed in ref.~\reff{15}, this is
obtained by
expressing the electromagnetic longitudinal and transverse
responses and, as well, their weak neutral current analogs in
terms of a single relativistic
{\it scaling} variable~\reff{16}. Moreover we use an approximation to the
relativistic electromagnetic and weak neutral current vertices,
 which has proven to be quite accurate. As we shall see,
this result is of relevance because the pionic effects in
these observables are felt up to quite large momentum transfers
($\sim 1$ GeV/c) where relativity cannot be ignored.

The items elaborated upon in the present paper are treated in the following
order: in sect.~2 we deal with the self--energy diagram both in
first--order perturbation theory and in the HF
scheme. In sect.~3 we consider the exchange diagram, again both in
first-- and infinite--order perturbation theory, where the latter is
explored within the framework of continued fractions.
In first--order, both forward-- and backward--amplitudes are included in the
 analysis.
In sect.~4 we calculate the electromagnetic and the weak neutral current
responses, focusing in particular on the large enhancement found for
the weak neutral current longitudinal response when correlations are included,
compared with the results found with the free RFG model where a delicate
cancellation occurs.
 In the same section we also address the question
of the evolution with $q$ of both the correlation and MEC contributions.
In sect.~5 we study the $\Lambda_\pi$-- and $k_F$--dependences of our
results. In the concluding section
we summarize our results as they stand at present in treating pionic
effects and their impact on quasielastic nuclear response functions.

\bigskip\bigskip\bigskip
\beginsection 2. Self--Energy Contributions

In this section we explore the contribution of the self--energy diagrams to
the longitudinal and transverse electromagnetic  responses of an infinite,
homogeneous nuclear system with an equal number of protons and neutrons
($Z=N=A/2$).
The corresponding (vector) weak neutral
current responses are considered in sect.~4,
whereas the axial--vector weak
neutral current response is studied in a companion paper~\reff{18}.
We shall confine ourselves to a consideration of
the one--particle--one--hole (1p--1h) sector of the excitation spectrum
of the RFG and, as already stated in sect.~1, in our scheme
only the pion mediates the interaction between the nucleons.

The nuclear responses may be expressed via the polarization
propagator. In first--order perturbation theory the pion yields two
self--energy contributions to the latter,
by dressing either the particle or the hole
propagation in the associated Goldstone diagrams. These are
displayed in fig.~1 and the corresponding analytic expressions for both the
longitudinal and transverse channels
have been derived in ref.~\reff{15}:
$$\eqalignno{& \Delta R_{L,T(s.e.)}^{corr}(q,\omega) = {\cal C}_{L,T}
{1\over m_N^4}\lim_{\alpha\to 0}{\partial\over\partial\alpha}\cr
&\quad\times\int
{d\kv\over{(2\pi)^3}}\theta(k_F-k)\theta(|\qv+\kv|-k_F)
\delta\biggl(\omega+\alpha-\kin\biggr)&\cr
&\quad\times
\int {d\pv\over{(2\pi)^3}}\theta(k_F-p)\Bigl\{\Gamma_\pi^2(\pv-\kv)
{(\pv-\kv)^2\over{(\pv-\kv)^2+\muq}}-\Gamma_\pi^2(\qv+\kv-\pv)
{(\qv+\kv-\pv)^2\over{(\qv+\kv-\pv)^2+\muq}}\Bigr\}\cr&&(2.1a)\cr
&\phantom{\Delta R_{L(s.e.)}^{corr}(q,\omega)} = {\cal C}_{L,T}
{1\over8\pi^2\kappa}\Biggl\{\theta(\lambda_2-\hla)\theta(\hla-|\lambda_1|)\cr
&\qquad\qquad\times{1\over2\kappa}
\Bigl[\biggl({\eta_F\bar{\psi}_r}
\biggr){\cal G}\biggl({\eta_F\bar{\psi}_r}\biggr)-
\biggl(\eta_F\psi_r\biggr){\cal G}\biggl(\eta_F\psi_r\biggr) \Bigr]\cr
&\qquad\qquad -\theta(\lambda_2-\hla)\theta(\hla-\lambda_1){\cal G}
\biggl(\sqrt{\eta_F^2+{4\hla}}
\biggr)+\theta(-\lambda_1-\hla){\cal G}\biggl(\sqrt{\eta_F^2-4\hla}
\biggr)\Biggr\},&(2.1b)\cr}$$
where $\xia=3\pi^2 A$, $\lm_2=\kappa^2+\eta_F\kappa$,
$\lm_1=\kappa^2-\eta_F\kappa$, $\hla=\lm(\lm+1)$, $\bar{\psi}_r=
 \psi_r+2\kappa/\eta_F$,
$$\eqalignno{
{\cal C}_L &= f_L^2(Q^2) {6\xia m_N^4\over{\kf^3}}\cou =
              f_L^2(Q^2) {6\xia m_N\over{\eta_F^3}}\cou, &(2.2a)\cr
{\cal C}_T &= f_T^2(Q^2) {6\xia m_N^4\over{\kf^3}}{q^2\over4m_N^2}
                          (\muv^2+\mus^2)\cou =
              f_T^2(Q^2) {6\xia m_N\over{\eta_F^3}}\kappa^2
                          (\muv^2+\mus^2)\cou\cr& &(2.2b)\cr}$$
and
$$\eqalignno{&{\cal G}(\z)= {1\over{4\pi^2}}\Bigl\{\mpi^3\Bigl[
\tan^{-1}({\eta_F+\z\over\mpi})+\tan^{-1}({\eta_F-\z\over\mpi})\Bigr]&\cr
&\quad -{\mpi^2\over{4\z}}(\eta_F^2+\mpi^2-\z^2)\ln\Bigl
\vert{(\eta_F+\z)^2+\mpi^2\over{(\eta_F-\z)^2+\mpi^2}}\Bigr\vert&\cr
&\quad +{\lam\over 2}(\lam^2-3\mpi^2)\Bigl[
\tan^{-1}({\eta_F+\z\over\lam})+\tan^{-1}({\eta_F-\z\over\lam})\Bigr]&\cr
&\quad+{1\over{4\z}}[\mpi^2(\eta_F^2-\z^2)-\lam^2(\lam^2-2\mpi^2)]
\ln\Bigl\vert{(\eta_F+\z)^2+\lam^2
\over{(\eta_F-\z)^2+\lam^2}}\Bigr\vert\Bigr\}.
&(2.3)\cr}$$
Here the contribution associated with the convective nucleon current has been
neglected, since it represents only a tiny fraction (a few percent, at most) of
the total transverse response for momentum transfers exceeding about 300 MeV/c.
Note that in (2.1) the derivative and the limit with respect to the variable
$\alpha$ define the prescription one should follow in order to deal with
the double pole characterizing the
first--order self--energy contribution to the nuclear responses.

In the above, along with the Fermi momentum $k_F$,
the three--momentum transfer $\qb$, the energy transfer $\om$,
and the spacelike four--momentum transfer
$Q^2=\om^2-q^2<0$, with $q=|\qb|$, the dimensionless variables of
ref.~\reff{15} have also been employed:
$$ \quad\kappa={q\over2m_N},\quad\lm={\om\over2m_N},\quad\eta_F={k_F\over m_N},
\quad\tau={|Q^2|\over4m_N^2},\quad\mu_\pi={m_\pi\over m_N}\quad {\rm and}\quad
\lm_\pi={\Lm_\pi\over m_N}. \eqno(2.4)$$
Moreover, $\muv=\mu_p-\mu_n$ and $\mus=\mu_p+\mu_n$ are the
isovector and the isoscalar magnetic moments of the nucleon, respectively.
To incorporate some aspects of
the $\pi$NN vertex, we employ the monopole form
$$\FFp(p)={\Lm^2_\pi-m_\pi^2\over\Lm^2_\pi+p^2}, \eqno(2.5)$$
which will be discussed at some length later on, especially in connection with
the problem of choosing an appropriate value for the phenomenological constant
$\Lm_\pi$.

Notice that in the energy--conserving $\delta$--function occurring
in our expressions for the self--energy contributions to $R_L$ and
$R_T$ we have replaced the
non--relativistic term $q^2/2m_N$ with the relativistic one $|Q^2|/2m_N$.
This can be shown~\reff{15} to be almost exactly equivalent to employing
relativistic kinematics and leads to
the scaling variable of the relativistic Fermi gas model~\reff{16}
which enters in (2.1):
$$\psi_r
={1\over\eta_F}\Big[{\lm(\lm+1)\over\kappa}-\kappa\Big] \eqno(2.6a)$$
(the last terms on the right--hand side of (2.1b) correspond to the
Pauli--blocked region, where no scaling occurs).

Had we kept the term $q^2/2m_N$ in the energy--conserving $\delta$--function,
then we would have obtained the same results but with the non--relativistic
scaling variable
$$\psi_{nr}={1\over\eta_F}\Big({\lm\over\kappa}-\kappa\Big) \eqno(2.6b)$$
rather than $\psi_r$. Actually, an additional factor $1+2\lm$ also appears in
the ``relativized'' responses: it expresses the Jacobian of the
transformation from the variable $\psi_{nr}$ to $\psi_r$ and has been
included in the multiplicative form factors (see below).

Another place where relativity is carefully accounted for in our approach
is in the electromagnetic vertices.
Indeed, the electromagnetic form factors $f_{L,T}(Q^2)$ presently employed are
given by the following expressions:
$$\eqalignno{
  f_L^2(Q^2) &= (1+2\lm)\left\{{1\over1+\tau}[G_E^{m_t}(\tau)]^2+
    \tau[G_M^{m_t}(\tau)]^2{\eta_F^2\over2}
    (1-\psi_r^2)\right\}&(2.7a)\cr
  f_T^2(Q^2) &= (1+2\lm){2\tau\over(\veps+\lm)^2}[G_M^{m_t}(\tau)]^2,
                        &(2.7b)\cr}$$
with
$$\eqalignno{
  G_E^{m_t}(\tau) &\equiv (\ums+m_t)G_{E_p}(\tau)+(\ums-m_t)G_{E_n}(\tau)
 &(2.8a)\cr
  G_M^{m_t}(\tau) &\equiv (\ums+m_t)G_{M_p}(\tau)+(\ums-m_t)G_{M_n}(\tau),
 &(2.8b)\cr}$$
$G_{E_{p,n}}$ and $G_{M_{p,n}}$ being the Sachs form factors of the
proton and neutron and $m_t$ labeling the isospin projection.
The results in (2.1) are actually obtained by adding the proton
response ($m_t=1/2$) multiplied by $Z$
to the neutron response ($m_t=-1/2$) multiplied by $N$.
Formulae (2.7) and (2.8) are deduced in ref.~\reff{15} and there shown to
provide
an accurate representation of the RFG responses.
We recall that the need for a fully--relativistic treatment of the
$\gm$NN vertex arises from the existence in the problem of the
non--relativistic reduction of the nuclear responses of a second scale
involving $\kappa$ and the squares of
the nucleon magnetic moments~\reff{16}, beyond the usual one set by
$\kappa$ alone.

A final comment on the expressions (2.1b), already discussed
in ref.~\reff{15}, relates to the domain in the ($q$,$\om$)--plane
where $\Delta R_{L(s.e.)}^{corr}$ and $\Delta R_{T(s.e.)}^{corr}$ are nonzero.
It is the same as that of the RFG, although the self--energy contributions do
not vanish
on the boundaries of the response region and are discontinuous across the
boundary dividing the Pauli--blocked region from the non--blocked one
(see the case at $q=300$ MeV/c below in fig.~2).
This is connected with the fact
that the derivative of the RFG response does not vanish on the boundaries
themselves and is discontinuous as well when one crosses into the
Pauli--blocked region.

As already mentioned, in the present approach we treat the pionic lines
of fig.~1 at the static level and hence they correspond to the
well--known one--pion--exchange potential (OPEP):
$$\eqalign{
  V(q) &= -{f_\pi^2\over m_\pi^2}\, \taub_1\cdot\taub_2 \, \sgb_1\cdot\qb
          \sgb_2\cdot\qb\,{1\over q^2+ m_\pi^2}\cr
       &= -{1\over3}{f_\pi^2\over m_\pi^2}\, \taub_1\cdot\taub_2
          [\sgb_1\cdot\sgb_2+S_{12}(\qbh)]\left(1-{m_\pi^2\over q^2+m_\pi^2}
          \right).\cr}\eqno(2.9)$$
In (2.9) one easily recognizes the central contact and momentum--dependent
pieces of the interaction in addition to the tensor
contribution. By performing spin traces (see Appendix A),
it is then an easy matter to show that {\it the tensor force
$S_{12}$ never contributes to the self--energy diagrams of fig.~1, i.e.,
in either longitudinal or transverse channels}.

It is equally easy to show, by the same token, that the longitudinal
isoscalar ($\tau=0$) and the isovector ($\tau=1$) self--energy
contributions are
identical, but for the difference between the isoscalar and the isovector form
factors $f_L^{\tau=0}$ and $f_L^{\tau=1}$.
This difference arises on the one hand because the neutron charge form
factor is nonzero (leading to a small effect). On the other hand,
the expressions (2.7a) contain terms which involve the proton and
neutron magnetic form factors and these
conspire to yield a much larger isovector than
isoscalar contribution (see below). Since these $G_M$--effects are
multiplied by $\eta_F^2\approx$ 0.04--0.08, they tend to be suppressed;
however, since they are multiplied by $\tau$, they provide increasingly
important contributions as the momentum transfer increases.  The usual
approach of including these (relativistic) corrections via the
Darwin--Foldy term already breaks down for $q<1$ GeV/c, as discussed in
ref.\reff{15}.

{}From (2.1) it clearly appears that the self--energy contribution actually
results from the
cancellation of two pieces associated, respectively, with the dressing of
the hole and particle propagation: the larger term is the latter
(at least
for $\Lambda_\pi$ not too small and $q$ not too large, see sect.~4 and fig.~2).
Therefore,
$\Delta R_{(s.e.)}^{corr}$ depletes the nuclear response at
low--$\omega$, while enhancing it at high--$\omega$,
as it should in order to obey the sum rule requirement~\reff{19}.
Worth noticing is that the contribution of the contact ($\delta$) term of
the OPEP cancels in (2.1), but for the momentum--dependence introduced by
the vertex function $\Gamma_\pi$. When the latter is taken into account,
 as in our case, then an additional cancellation between
the $\delta$-- and momentum--dependent contributions occurs.
What remains is then a rather modest
contribution (negative at small and positive at large $\om$, again for not too
small $\Lambda_\pi$ and not too large $q$), that rapidly decreases
as $q$ increases.
This is illustrated in fig.~2 for $q=300$, 500 and 1000 MeV/c in the
longitudinal channel. For sake of illustration here and in the following we
take $k_F=225$ MeV/c (which roughly corresponds to the case of
$^{12}$C)
and $\Lambda_\pi=1300$ MeV, unless otherwise specified.

A similar situation also holds in the transverse channel, but for
the well--known factor $Z\mu_p^2+N\mu_n^2={1\over 2}A(\mu_p^2 + \mu_n^2)=
{1\over 4}A(\mus^2 + \muv^2)$, which greatly suppresses the isoscalar
contribution compared to the isovector contribution because of the fact
that $(\mus/\muv)^2\cong 0.035$.

An issue we wish to address in closing this section relates to the
self--energy diagrams of higher--order in perturbation theory. Do they play an
important role? To answer this question at least partially,
we may examine the longitudinal
electromagnetic response in the Hartree--Fock approximation (actually here
only the Fock approximation, since the Hartree contribution
vanishes in nuclear matter, although we shall continue to use the label
HF) and compare with the results displayed in fig.~2
to which we add the free response.
The HF longitudinal response reads~\reff{15}
$$\eqalignno{&R_L^{HF}(q,\omega)= {\xia\over{2(2\pi)^3 \eta_F^3\M}}
f^2_L(Q^2)\int d\betv\theta(\eta_F-\beta)\theta(|2\rv+\betv|-\eta_F)&\cr
&\qquad\times\delta\biggl\{\lambda-\kina-
[{\widetilde{\Sigma}}_{\pi}^{(1)}(|\rv+\betv|)-
{\widetilde{\Sigma}}_{\pi}^{(1)}(\beta)]\biggr\},
&(2.10)\cr}$$
where
$$\widetilde{\Sigma}_{\pi}^{(1)}(\bt)= 3\cou
{\M^2\over\eta_F^2}{\cal G}(\bt)\eqno(2.11)$$
is the pion self--energy.

In fig.~3 the longitudinal HF
response is displayed for $q=$300, 500 and 1000 MeV/c.
Note that this response appears in a region of the ($q$,$\om$)--plane
somewhat different from that where the RFG responses are nonzero,
notwithstanding that each order of perturbation
theory in the self--energy is actually nonzero precisely where the RFG is
nonzero. The effect, however, is so small (at most a few MeV) that it can only
be perceived at low--$q$, as seen in fig.~3. Remarkably, from this
figure one realizes that the first--order response
practically coincides with the HF result: the many--body framework apparently
quells the
strong interaction carried by the pion to the point of rendering quite accurate
a first--order perturbative treatment of the OPEP, at least as far as the
mean--field contributions are concerned.
Accordingly, the impact of the pion self--energy contribution
on the total charge response is rather mild and one
observes some hardening for $q\approx$ 300 MeV/c, which, however,
disappears as $q$ increases (see sect.~6).

\bigskip\bigskip\bigskip
\beginsection 3. Exchange contributions

In the 1p--1h sector of the nuclear response,
in addition to the
self--energy diagrams dealt with in the previous section, two further
classes of Goldstone diagrams
yield contributions in first--order perturbation theory: they are
displayed in fig.~4 and are commonly referred to as {\it exchange diagrams}.
In particular, in fig.~4a the 1p--1h state propagates forward in time
(as in the so--called Tamm--Dancoff approximation (TDA)),
whereas in fig.~4b a backward--going 1p--1h state occurs
(as when one incorporates ground--state correlations
in the RPA). The forward-- and
backward--going first--order perturbation theory contributions will be
denoted $F$ and $B$, respectively.
The expressions for the associated contributions to the electromagnetic charge
response read:
$$\eqalignno{
  &\Delta R_{L(exch,{\rm F})}^{corr}(q,\om) = f_L^2(Q^2)\xi_A{2\over k_F^3}
    {f^2_\pi\over m_\pi^2} \cr
  &\qquad\times\int\! {d\kb_1\over(2\pi)^3}{d\kb_2\over(2\pi)^3}
    \theta(|\kb_1+\qb|-k_F)\theta(k_F-k_1)\theta(|\kb_2+\qb|-k_F)
    \theta(k_F-k_2) \cr
  &\qquad\quad\times\Gamma_\pi(\kb_1+\qb)\Gamma_\pi(\kb_1)
    {\dl(\om-{|Q^2|/2m_N}-{\qb\cdot\kb_1/ m_N})\over
    \om-{|Q^2|/2m_N}-{\qb\cdot\kb_2/ m_N}}
    {(\kb_1-\kb_2)^2\over m_\pi^2+(\kb_1-\kb_2)^2}&(3.1a) \cr
  &\phantom{\Delta R_{L(exch,{\rm F})}^{corr}(q,\om)}
    =f_L^2(Q^2){1\over128\pi^4}{\xi_A\over\eta_F\kappa^2}{f_\pi^2\over m_N}\cr
  &\qquad\times\left\{\theta(\eta_F-\kappa)\int_{k_1^<}^1\!dk_1
    \int_{|2\kappa/\eta_F-1|}^1\!dk_2\big[\Nc_L(k_1,k_2;k_2,\psi_r)-
    \Nc_L(k_1,k_2;t,\psi_r)\big]\right. \cr
  &\qquad\quad+\left.\theta(2\kappa-\eta_F)\int_{k_1^<}^1\!dk_1
    \int_0^{k_2^>}\!dk_2\big[\Nc_L(k_1,k_2;k_2,\psi_r)-
    \Nc_L(k_1,k_2;-k_2,\psi_r)\big]\right\} \cr&&(3.1b)\cr}$$
and
$$\eqalignno{
  &\Delta R_{L(exch,{\rm B})}^{corr}(q,\om) =-f_L^2(Q^2)\xi_A{2\over k_F^3}
    {f^2_\pi\over m_\pi^2} \cr
  &\qquad\times\int\! {d\kb_1\over(2\pi)^3}{d\kb_2\over(2\pi)^3}
    \theta(|\kb_1+\qb|-k_F)\theta(k_F-k_1)\theta(k_F-|\kb_2+\qb|)
    \theta(k_2-k_F) \cr
  &\qquad\quad\times\Gamma_{\pi}(\kb_1+\qb)\Gamma_{\pi}(\kb_1)
    {\dl(\om-{|Q^2|/2m_N}-{\qb\cdot\kb_1/ m_N})\over
    \om-{|Q^2|/2m_N}-{\qb\cdot\kb_2/ m_N}}
    {(\kb_1-\kb_2)^2\over m_\pi^2+(\kb_1-\kb_2)^2} &(3.2a)\cr
  &\phantom{\Delta R_{L(exch,{\rm B})}^{corr}(q,\om)}
  =f_L^2(Q^2){1\over128\pi^4}{\xi_A\over\eta_F\kappa^2}{f_\pi^2\over m_N}\cr
  &\qquad\times\left\{\theta(\eta_F-\kappa)
    \int_{\bar{k}_1^<}^{\bar{k}_1^>}\!dk_1
    \int_{|2\kappa/\eta_F-1|}^1\!dk_2\big[\Nc_L(k_1,k_2;k_2,-\psibar)-
    \Nc_L(k_1,k_2;t,-\psibar)\big]\right. \cr
  &\qquad\quad+\left.\theta(2\kappa-\eta_F)
    \int_{\bar{k}_1^<}^{\bar{k}_1^>}\!dk_1
    \int_0^{k_2^>}\!dk_2\big[\Nc_L(k_1,k_2;k_2,-\psibar)-
    \Nc_L(k_1,k_2;-k_2,-\psibar)\big]\right\}, \cr&&(3.2b)\cr}$$
respectively.

In the transverse channel we have instead
$$\eqalignno{
  &\Delta R_{T(exch,{\rm F})}^{corr}(q,\om) = f_T^2(Q^2)\xi_A{2\over k_F^3}
    {f^2_\pi\over m_\pi^2} {1\over4m_N^2}(\muv^2-3\mus^2)\cr
  &\qquad\times\int\! {d\kb_1\over(2\pi)^3}{d\kb_2\over(2\pi)^3}
    \theta(|\kb_1+\qb|-k_F)\theta(k_F-k_1)\theta(|\kb_2+\qb|-k_F)
    \theta(k_F-k_2) \cr
  &\qquad\quad\times\Gamma_\pi(\kb_1+\qb)\Gamma_\pi(\kb_1)
    {\dl(\om-{|Q^2|/2m_N}-{\qb\cdot\kb_1/ m_N})\over
    \om-{|Q^2|/2m_N}-{\qb\cdot\kb_2/ m_N}}
    {[\qb\cdot(\kb_1-\kb_2)]^2\over m_\pi^2+(\kb_1-\kb_2)^2} &(3.3a)\cr
  &\phantom{\Delta R_{T(exch,{\rm F})}^{corr}(q,\om)}
  =f_T^2(Q^2){1\over128\pi^4}{\xi_A\over\eta_F}{f_\pi^2\over m_N}
     (\muv^2-3\mus^2)\cr
  &\qquad\times\left\{\theta(\eta_F-\kappa)\int_{k_1^<}^1\!dk_1
    \int_{|2\kappa/\eta_F-1|}^1\!dk_2\big[\Nc_T(k_1,k_2;k_2,\psi_r)-
    \Nc_T(k_1,k_2;t,\psi_r)\big]\right. \cr
  &\qquad\quad\left.\phantom{\int_|^1}
    +\theta(2\kappa-\eta_F)\,\psi_r\big[\Hc_T(1,q_<)-
    \Hc_T(k_1^<,q_<)\big] \right\} &(3.3b)\cr}$$
and
$$\eqalignno{
  &\Delta R_{T(exch,{\rm B})}^{corr}(q,\om) = -f_T^2(Q^2)\xi_A{2\over k_F^3}
    {f^2_\pi\over m_\pi^2} {1\over4m_N^2}(\muv^2-3\mus^2)\cr
  &\qquad\times\int\! {d\kb_1\over(2\pi)^3}{d\kb_2\over(2\pi)^3}
    \theta(|\kb_1+\qb|-k_F)\theta(k_F-k_1)\theta(k_F-|\kb_2+\qb|)
    \theta(k_2-k_F) \cr
  &\qquad\quad\times\Gamma_\pi(\kb_1+\qb)\Gamma_\pi(\kb_1)
    {\dl(\om-{|Q^2|/2m_N}-{\qb\cdot\kb_1/ m_N})\over
    \om-{|Q^2|/2m_N}-{\qb\cdot\kb_2/ m_N}}
    {[\qb\cdot(\kb_1-\kb_2)]^2\over m_\pi^2+(\kb_1-\kb_2)^2} &(3.4a)\cr
  &\phantom{\Delta R_{T(exch,{\rm B})}^{corr}(q,\om)}
  =f_T^2(Q^2){1\over128\pi^4}{\xi_A\over\eta_F}{f_\pi^2\over m_N}
     (\muv^2-3\mus^2)\cr
  &\qquad\times\left\{\theta(\eta_F-\kappa)
    \int_{\bar{k}_1^<}^{\bar{k}_1^>}\!dk_1
    \int_{|2\kappa/\eta_F-1|}^1\!dk_2\big[-\Nc_T(-k_1,k_2;k_2,-\psibar)+
    \Nc_T(-k_1,k_2;t,-\psibar)\big]\right. \cr
  &\qquad\quad\left.\phantom{\int_|^|}-\theta(2\kappa-\eta_F)\,
    \psibar\big[\Hc_T(\bar{k}_1^>,q_<)-
    \Hc_T(\bar{k}_1^<,q_<)\big] \right\}. &(3.4b)\cr}$$
In (3.1-4) the following quantities have been introduced:
$$
\eqalignno{
  t &= {\eta_F^2-4\kappa^2-k_2^2/m_N^2\over4\eta_F\kappa}\cr
\noalign{\medskip}
  k_1^< &= \left\{\matrix{
  \sqrt{1-(2\kappa/\eta_F)^2-4\kappa\psi_r/\eta_F}, &\psi_r<1-2\kappa/\eta_F\cr
  |\psi_r|, &\psi_r>1-2\kappa/\eta_F\cr}\right.\cr
\noalign{\medskip}
  k_2^> &= \left\{\matrix{2\kappa/\eta_F-1, &\kappa<\eta_F\cr
                         1, &\kappa>\eta_F\cr}\right.\cr
\noalign{\medskip}
  \bar{k}_1^< &= \left\{\matrix{
    1, &\psi_r<1-2\kappa/\eta_F\;\rightarrow\;\bar{\psi}_r<1\cr
    \bar{\psi}_r, &\psi_r>1-2\kappa/\eta_F\;\rightarrow\;\bar{\psi}_r>1\cr}
  \right.  &(3.5) \cr
\noalign{\medskip}
  \bar{k}_1^> &= \sqrt{1+(2\kappa/\eta_F)^2+4\kappa\psi_r/\eta_F}
               = \sqrt{1-(2\kappa/\eta_F)^2+4\kappa\bar{\psi}_r/\eta_F}\cr
\noalign{\medskip}
  q_< &=\left\{\matrix{2\kappa/\eta_F, &\kappa<\eta_F\cr
                       2, &\kappa>\eta_F\cr}\right.\cr}$$
whereas the functions $\Nc_{L,T}$ and $\Hc_T$ are defined in Appendix~B.
Note that while we have been able to express the
exchange contribution to the transverse response analytically, at least when
 $\kappa>\eta_F$, this
turned out not to be possible in the case of the longitudinal channel.

A few remarks should be made concerning the above expressions.
First, notice that all of the exchange contributions are
 different from zero in the
response region of the RFG, {\it vanishing}, however, on its boundaries (at
variance with the self--energy case).
Second, we note that the exchange contributions
can be split, as in the self--energy case,
into isoscalar and isovector components. By carrying out
the pertinent traces over the Pauli matrices (see Appendix A), in this case
it is found that
in the $\tau=0$ channel the pion exchange correlations are
 {\it three times stronger}
than in
$\tau=1$ case, being moreover of opposite sign in the two isospin channels.
Finally, performing a similar analysis in spin space (see Appendix A),
one finds that the tensor interaction $S_{12}$ {\it does not contribute} to the
charge response, but only to the spin response.

The results of our calculation of the exchange contribution to the nuclear
responses are displayed in fig.~5, where both
the forward-- and backward--going exchange contributions to the charge
 response are displayed for
$q=$300, 500 and 1000 MeV/c, and in fig.~6, where the same is done for the spin
response.
A feature common to all of the results we present here is immediately apparent:
strength is removed by the pionic exchange correlations from the
low--energy side of the nuclear responses and added to the high--energy side.
This {\it hardening} effect~\reff{20} goes in parallel with the one induced by
the self--energy correlations discussed in sect.~2, which however was
somewhat milder. As a consequence both charge and spin responses turn out
to be hardened with respect to the corresponding free RFG responses.

The hardening of the charge response
arises from the strong {\it repulsive}
character of the exchange isoscalar pionic correlations, which are
overwhelmingly carried by the $\dl$--component of the OPEP (the {\it
attractive} isovector correlations for zero--range contributions are
three times weaker, as mentioned above). The finite--range central
interaction of the OPEP, while attractive is not very effective and
only modestly reduces the impact of the $\dl$--force on the charge response.
Also worth mentioning is that the role of $\dl$--contributions
is clearly apparent in the $F$--term  and becomes dominant in the
$B$--term.

In the spin channel the isoscalar correlations are dramatically suppressed
relative to the isovector correlations
by the factor $(\mus/\muv)^2\cong 0.035$. In contrast to
the situation for the charge case, the isovector central
correlations are now of a
{\it repulsive} character (see Appendix A for the derivation of the
related spinology), leading again to a hardening of the spin
response due to the action of the $\dl$--force. The latter, however,
is counteracted here not only by the attractive momentum--dependent
central interaction, but by the tensor force as well. Accordingly, the
amount of hardening of the spin response is somewhat moderated with
respect to the charge case.

Interestingly, in the spin channel the $\dl$--dominance is restricted only to
the $F$--diagram, since the largest contribution to the $B$--diagram is
in fact provided by the tensor force --- and the latter is known to be quite
efficient in providing ground--state correlations.
This explains why the contribution from the $B$--diagram survives in the spin
channel up to larger momenta than in the charge channel.
With regard to the momentum--dependence of the exchange contributions to the
nuclear responses, one should also notice that the $B$--term falls faster
than the $F$--term as the momentum transfer increases because, as seen from
(3.1b), (3.2b), (3.3b) and (3.4b), the integrals involved in the former
span a higher momentum range than those in the latter. As a
consequence, the backward--going responses are cut--off more rapidly
by $\Gamma_{\pi}$ than are the forward--going ones.

In summary, from the above analysis the following observations can be
made:

a) the exchange diagrams of fig.~4 are the ones that contribute most
strongly to the nuclear responses as first--order perturbations in the
1p--1h sector of the RFG;

b) their magnitude and sign is to a large extent set by the $\dl$--component
of the OPEP.

\noindent This last finding will be
addressed further in sect.~5 in connection with the discussion of the global
gauge invariance of our theory.

A further aspect of our results worth noting is that the sum--rule
 requirement is indeed fulfilled
by the $F$--contribution, whereas it is violated by the $B$--term.~\reff{21}
Indeed, the former is characterized by a node in the response function
(when expressed as a function of $\omega$)
located at the peak position of the RFG response. Here the very tiny
discrepancies seen
can be ascribed to the present approximate, although very good,
treatment of the relativistic kinematics. In contrast,
the latter never changes sign, always remaining negative.
In this connection one sees that the hardening of the nuclear
 responses previously discussed clearly follows from the way the sum rule is
fulfilled by the $F$--diagram. In addition a significant contribution
to the hardening also comes from the $B$--diagram. In fact, because the
(negative) $B$--term
obeys the energy--weighted sum rule, it then reaches its minimum somewhat
before the peak of the free response.

To conclude this section we now explore the role played by
exchange diagrams of
order higher than one in the nuclear response. We do so for the charge
longitudinal channel in the TDA, which amounts to accounting for all terms
in the series shown in fig.~7. To achieve this, we employ the method of
continued fractions~\reff{22}: in this framework the polarization propagator
becomes
$$\Pi^{\rm CF}_{L,\tau}(\kappa,\lm)={\Pi^0(\kappa,\lm)\over
          1+\Pi^{(1)}_{L,\tau}(\kappa,\lm)\big/\Pi^0(\kappa,\lm)+...}\;,
  \eqno(3.6)$$
where $\Pi^{(1)}_{L,\tau}$ is the longitudinal first--order exchange
contribution
in the isospin channel $\tau$ and $\Pi^0$ the RFG propagator. The imaginary
part of $\Pi^{(1)}_{L,\tau}$ is proportional to (3.1), whereas the real part
is given in Appendix B.

We limit our attention to the first one of the infinite set of continued
fractions embodied in (3.6), recalling that in the limit of a
zero--range force the
first iteration of (3.6) actually corresponds to the exact TDA solution.
As previously discussed, since the $\dl$--piece of the interaction
plays the largest role in the charge response, we believe
that the results displayed in fig.~8 are quite representative of the true
TDA solution (of course, one should also take into account that
in coordinate space $\Gamma_\pi$
actually smears out the $\dl$--function).

In order to characterize the results obtained here, in fig.~8 two
cases are shown:
i) the longitudinal TDA exchange response with a pure $\dl$--interaction
together with
$\Gamma_\pi$ and ii) adding to this the momentum--dependent forces as
well. The latter, which corresponds to a weaker interaction,
shows that the
first--order result seems to be quite accurate even at
$q=$500 MeV/c. However, the
degree of accuracy appears to be lower than in the corresponding self--energy
case, in accord with the finding that the action of the pion is stronger in
the exchange channel than in the self--energy channel.

\bigskip\bigskip\bigskip
\beginsection 4. Electromagnetic and weak neutral current responses

In the previous two sections we have set up all the ingredients needed
to calculate not only the electromagnetic responses, but also the weak neutral
current ones~\reff{23-27} in first-- (and as well in infinite--) order
perturbation
theory, within a framework where only the pion--induced correlations are
taken into account.
Indeed, by splitting the electromagnetic responses into their isospace
components according to
$$\eqalignno{
  R_L^{\rm em} &= R_L^{\rm em}(\tau=0) + R_L^{\rm em}(\tau=1) &(4.1a)\cr
  R_T^{\rm em} &= R_T^{\rm em}(\tau=0) + R_T^{\rm em}(\tau=1) &(4.1b)\cr}$$
and exploiting the conservation of the vector current
 (CVC)~\reff{26}, one easily obtains the weak
neutral current longitudinal and transverse responses (as mentioned earlier,
the axial--vector one is
discussed in a companion paper~\reff{18}). They read
$$\eqalignno{
  R^L_{AV} &= a_A\left[\bt^{(0)}_V R_L^{\rm em}(\tau=0) +
\bt^{(1)}_V R_L^{\rm em}(\tau=1)\right]     &(4.2a)\cr
  R^T_{AV} &= a_A\left[\bt^{(0)}_V R_T^{\rm em}(\tau=0) +
\bt^{(1)}_V R_T^{\rm em}(\tau=1)\right],    &(4.2b)\cr}$$
where $a_A=-1$ and the isoscalar and isovector weak neutral current couplings,
$\bt^{(0)}_V$ and $\bt^{(1)}_V$, respectively, are fixed at tree--level in the
standard model to the values
$$\eqalignno{
  \bt^{(0)}_V &= -2\sin^2\th_W \approx-0.454 &(4.3a)\cr
\noalign{and}
  \bt^{(1)}_V &= 1-2\sin^2\th_W \approx0.546\,, &(4.3b)\cr}$$
using $\sin^2\th_W\approx0.227$ for the Weinberg angle.

To the contributions stemming from the correlations, those arising from the
MEC should then be added. These have been thoroughly
discussed within the context of 1p--1h excitations
in ref.~\reff{15}, where their contribution to the electromagnetic
responses has been calculated including terms of order up to $\kappa^2$ in
a non--relativistic reduction, where, in
addition to spatial components, the time component of
the MEC has also been kept in the evaluation.
The ensuing violation of the continuity equation thus introduced has
been checked in ref.~\reff{15} and found to be quite small.
The associated expressions in the weak neutral current sector are most easily
obtained by recalling that the MEC for leading--order $\pi$--exchange
are purely isovector in nature. Accordingly one simply obtains
$$\eqalignno{
  R^L_{AV}({\rm MEC}) &= a_A \bt^{(1)}_V R_L^{\rm em}({\rm MEC}) &(4.4a)\cr
  R^T_{AV}({\rm MEC}) &= a_A \bt^{(1)}_V R_T^{\rm em}({\rm MEC}) &(4.4b)\,.
  \cr}$$

In figs.~9 and 10 we now display the electromagnetic responses (longitudinal
and transverse, respectively) at $q=$300, 500 and 1000 MeV/c.
We observe that
in both channels a hardening of the responses occurs,
particularly at low--$q$, essentially
induced by the $\dl$--force via the exchange diagram, as discussed at length in
the previous section. Also in accord with the considerations
developed there the shift in the maximum of the response is seen to be larger
in the charge channel than in the spin one.
As the momentum transfer increases, the amount of hardening
decreases, since the pion--induced correlations weaken for large $q$. This
behaviour partly reflects the role of the hadronic vertex function
$\Gamma_\pi$ in cutting out the high--momentum contributions to the response
functions and partly reflects the
constraints arising from the many--body system  itself, which appear through
the mismatch between the single--particle wave functions of the particle--hole
pairs involved in the intermediate states of the polarization propagators
considered here.

Clearly, a unique way to assess the impact of correlations on the response
functions does not exist.
Among several different possibilities, we have chosen to focus on
the following two quantities:
i) the shift $\Delta\om_{\rm QEP}$ of the peak position of the response with
respect to the corresponding quantity in the RFG, which is displayed versus
$q$ in fig.~11 for both the longitudinal and transverse channels;
ii) the correlated sum rule, {\it i.e.\/} the integral of the
nuclear responses over $\omega$, divided, in the longitudinal and transverse
channels, by the functions
$$C_{L,T}(\tau,\psi_r;\eta_F)={{\cal N}\over 4m_N\kappa}\times
 \left\{\eqalign{ &{\kappa^2\over\tau}\left[(1+\tau)
W_2(\tau)-W_1(\tau)+W_2(\tau)\Delta
\right] \cr
         &2W_1(\tau)+W_2(\tau)\Delta \cr}\right.
\qquad \eqalign{&\vphantom{G^2_E}{\rm for}\; L\cr
          &\vphantom{{1\over 2}}{\rm for}\; T\cr}
\eqno(4.5)$$
respectively.
As shown in ref.~\reff{16}, this optimally separates the nucleonic
physics from the many--body problem for the RFG. In (4.5)
$$\eqalignno{W_1(\tau)&=\tau G_M^2(\tau) &(4.6a)\cr
W_2(\tau)&={1\over 1 +\tau}\left(G_E^2(\tau)+\tau G_M^2(\tau)\right).
&(4.6b)\cr}$$
and
$$\Delta={\tau\over\kappa^2}(1-\psi_r^2)\xi_F\left\{\kappa\sqrt{1+{1\over\tau}}
 +{1\over 3}(1-\psi_r^2)\xi_F\right\},
\eqno(4.7)$$
$\psi_r$  being the scaling variable (2.6a) and  $\xi_F=\sqrt{1+\eta_F^2}
-1$. We recall that with these definitions we actually integrate the part
of the responses which scales. So it should not be surprising that the
asymptotic value, namely one, is reached for momenta much larger than $2k_F$.
The normalized sum rules are reported versus $q$ in figs.~12a and 12b,
again for both channels. We recall that the sum rules
reflect only the roles of the backward--going correlations and of the MEC.

In accord with the discussion of sect.~3, from these figures one first
infers that the correlations persist up to
larger momenta in the longitudinal channel than in the transverse
 one; secondly, one sees that the MEC contribution, although not
large, is significant, rather constant with $q$ and acts in a way that
depletes the sum rule. Finally inspection
of the responses shows that in the longitudinal channel the $B$--terms decrease
in magnitude with $q$ more rapidly than the correlations stemming from the
$F$--diagram, whereas in the transverse channel the opposite occurs ---
a clear signature of the action of the tensor force.

Turning now to a discussion of the weak neutral
current responses, we first consider the charge
channel. To understand the nature of the problem, we
display in fig.~13 the longitudinal  nuclear response at $q=$300 MeV/c
separated into $\tau=0$ and $\tau=1$ channels. There we see that the action
of the pion is very important in hardening the
isoscalar channel, while rather gentle in softening the isovector one.
Note that the effect of the
longitudinal electromagnetic form factors in contributing to this imbalance is
negligible at this low momentum transfer. When we combine the two isospin
contributions with weighting factors $\bt^{(0)}_V$ and $\bt^{(1)}_V$
according to (4.2a) to get the longitudinal weak neutral
current response we obtain
the result shown in fig.~14. While the magnitude of free response is
quite small, the pion--correlated one is dramatically modified.  The origin
of this effect was discussed previously in ref.~\reff{26} (see also
refs.~\reff{27,24}), where it was noted
that the delicate cancellation that leads to the suppression of $R^L_{AV}$
in the RFG model is broken when the isoscalar and isovector channels are
correlated differently. This heightened sensitivity to isospin correlation
effects opens a new window on nuclear physics in the quasielastic region,
possibly offering the unique
possibility of separating the two isospin channels, something that is
impossible to achieve with parity--conserving electron scattering alone.
It can, however, be brought to
light using parity--violating electron scattering, as discussed in
ref.~\reff{26}, where it is shown that isospin correlations have
dramatic consequences for the forward--angle parity--violating
asymmetry over a range of momentum transfers persisting up
to $q\approx500$ MeV/c.

In concluding this section, we return to touch on the transverse weak
neutral current
response $R^T_{AV}$: here the factor $\mus^2$ suppresses the isoscalar
contributions and consequently no sensitive cancellation occurs in the
free RFG model to be broken by isospin correlations, in contrast to the case
of $R^L_{AV}$. Accordingly, the prediction for
$R^T_{AV}$ is very simple: it is approximately half ($\bt_V^{(1)}=0.546$)
of the corresponding electromagnetic response (of course in (4.2a,b) the
electromagnetic and weak neutral current coupling constants are not included).
These observations were critical in our earlier work~\reff{26} where the
insensitivity of the transverse response functions to isospin correlation
effects allowed us to suggest backward--angle parity--violating quasielastic
electron scattering as a tool to probe the single--nucleon form factors
themselves.  We shall return to quantify the roles played by pionic
effects in this situation in ref.~\reff{18}.

\bigskip\bigskip\bigskip
\beginsection 5. $\Lambda_\pi$ and $k_F$ dependence of the nuclear responses

In this section we explore the significance of the roles played by the two
parameters on which our calculation of the nuclear responses depends, namely
$\Lambda_\pi$ and $k_F$. We start by considering $\Lambda_\pi$, which
incorporates some aspects of the short--range physics affecting the pionic
correlations.
To get a feeling for the distances involved in our problem, we consider the
Fourier transform of $\Gamma_\pi$ (see (2.5)), namely
$$\Gamma_\pi(r)=(\Lambda_\pi^2-m_\pi^2){{\rm e}^{-\Lambda_\pi r}\over4\pi r}.
  \eqno(5.1)$$
The corresponding rms radius of the interaction region
$$\sqrt{<r^2>} = \Bigl[\int\!d\rb\,r^2\Gamma_\pi(r)\Bigr]^{1/2}=
  {\sqrt{6}\over{\Lambda_\pi}} \sqrt{1-
({m_\pi/\Lambda_\pi})^2}\ , \eqno(5.2)$$
is then easily obtained. Typical values of $\Lambda_\pi$ are of order
1--2 GeV and hence $(m_\pi/\Lambda_\pi)^2\sim$ 0.005--0.02, namely, only
a small correction in (5.2).  Accordingly the rms radius above is given to
a good approximation by $\sqrt{6}/\Lambda_\pi$: for the value $\Lambda_\pi=$
1300 MeV used in most of this work, this yields an rms radius of 0.37 fm.

On the one hand, as seen in fig.~15 the sensitivity of the exchange
contribution to variations in $\Lambda_\pi$ appears to be quite mild.
For sake of illustration we limit ourselves
here and in the following to consideration only of the $q=500$~MeV/c case
and only of the longitudinal channel. The self--energy contribution, on the
other hand, is more sensitive to changes in $\Lambda_\pi$, even reversing its
behaviour in the energy variable for
$\Lambda_\pi\approx800$ MeV, reflecting the weakening of the particle
self--energy with respect to that for the hole (fig.~15).

Somewhat in between the two situations discussed above is the
behaviour versus $\Lambda_\pi$ of the MEC (fig.~16). We may conclude
that, owing to the major contribution to the nuclear
responses arising from the exchange term,
our pionic model is indeed affected,
but only mildly, by the
short--range physics embodied in the vertex function $\Gamma_\pi$ for
$\Lambda_\pi\rapp1.3$ GeV.

Clearly, all of the short--range physics cannot be represented solely by the
monopole form factor in (5.1).
An adequate treatment of the short--range physics (presuming we continue to
adopt a strategy of sticking to a mesonic
model of the nucleus) would require on the one hand the introduction of
heavier mesons, and on the other the calculation of additional contributions
({\it e.g.,\/} the ladder diagrams) beyond the self--energy
and exchange terms considered in the present
approach. An alternative option would of course be to abandon the
hadronic description of the short--range physics in favour of QCD degrees of
freedom, but this has not yet proven to be tractable in the strong--coupling
regime.

Next let us comment on how the continuity equation is affected by the
vertex function $\Gamma_\pi$. In this connection we recall that the
continuity equation in momentum space reads
$$i\left(\pb_1+\pb_2\right)\cdot\Jb_{\rm MEC}\left(\pb_1,\pb_2\right) =
\left(\taub_1\times\taub_2\right)^3\left[V(\pb_2)-V(\pb_1)\right],
\eqno(5.3)$$
where $\pb_1$ and $\pb_2=\qb-\pb_1$ are the momenta carried by the meson.
For pointlike nucleons (5.3) is fulfilled by the OPEP (2.9) and by
the longitudinal components of the pion--in--flight and contact currents:
$$\eqalign{
\left(\pb_1+\pb_2\right)\cdot\Jb_\pi\left(\pb_1,\pb_2\right)
&= -i{f_\pi^2\over m_\pi^2}\, \left(\taub_1\times\taub_2 \right)^3\cr
&\times\sgb_1\cdot\pb_1\sgb_2\cdot\pb_2
\left({1\over m_\pi^2+\pb_1^2}-{1\over m_\pi^2+\pb_2^2}\right)
\cr}\eqno(5.4a)$$
and
$$\eqalign{
&\left(\pb_1+\pb_2\right)
\cdot\Jb_{contact}\left(\pb_1,\pb_2\right)
= -i{f_\pi^2\over m_\pi^2}\, \left(\taub_1\times\taub_2 \right)^3\cr
&\times\left[
{\sgb_1\cdot\pb_1\sgb_2\cdot\pb_2\over m_\pi^2+\pb_2^2}+
{\sgb_1\cdot\pb_2\sgb_2\cdot\pb_2\over m_\pi^2+\pb_2^2}-
{\sgb_1\cdot\pb_1\sgb_2\cdot\pb_1\over m_\pi^2+\pb_1^2}-
{\sgb_1\cdot\pb_1\sgb_2\cdot\pb_2\over m_\pi^2+\pb_1^2}
\right]\ .\cr}\eqno(5.4b)$$

It is of significance that the $\dl$--piece of the OPEP does not
contribute to (5.3), since its commutator with the charge operator vanishes.
Accordingly, to keep or to drop the $\dl$--term in the potential
is of considerable relevance for the nuclear responses, but is
irrelevant as far as the global gauge invariance of the theory is concerned.
It becomes of importance, however, when the  vertex function $\Gamma_\pi$
is included. In fact,
in order to satisfy the continuity equation in such a case
new terms should be added to the pion--in--flight current, for example, the
following~\reff{17}
$$\eqalignno{
  \Jb_\pi\left(\pb_1,\pb_2\right) &=
  i {f_\pi^2\over m_\pi^2}(\taub_1\times\taub_2)^3(\pb_1-\pb_2)
  (\Lambda_\pi^2-m_\pi^2)^2\sgb_1\cdot\pb_1\,\sgb_2\cdot\pb_2 \cr
  &\quad\times\left\{{1\over(\Lambda_\pi^2+\pb_1^2)}\left[
  {1\over(\Lambda_\pi^2+\pb_1^2)(m_\pi^2+\pb_1^2)} +
  {1\over(m_\pi^2+\pb_1^2)(m_\pi^2+\pb_2^2)} \right.\right.\cr
  &\quad\qquad\left.\left.+
  {1\over(\Lambda_\pi^2+\pb_2^2)(m_\pi^2+\pb_2^2)}\right]
  {1\over(\Lambda_\pi^2+\pb_2^2)}\right\}. &(5.5)\cr}$$
In contrast, no modifications are necessary but for
the multiplicative factor $\Gamma_\pi^2$ occurring in
the contact (or pair) term,
this being the one contributed to by the $\dl$ (smeared) piece of
the OPEP.
The first and the third terms on the right--hand side of (5.5) may be viewed
as describing the coupling of the virtual photon to a fictitious particle
carrying the same quantum numbers as the pion, which thus underlies the
microscopic description of $\Gamma_\pi$.
We have numerically checked the importance of these additional terms
and found them quite small (about 10\% of
the pion--in--flight contribution at the peak of the latter in the charge
channel and about 20\% in the spin channel under the same conditions and so
yielding less than 1\% corrections to the total responses). This
occurs unless $\Lambda_\pi$ becomes
unreasonably small --- say a few hundred MeV, a range of values that not only
can hardly be accepted, but is also such as to render the overall pionic
corrections to the RFG responses quite negligible. It remains however to
be verified whether the above findings are still  valid in the 2p--2h
sector or when mesons heavier than the pion are brought into play.

As a general remark we note that, even in the presence of
the vertex function $\Gamma_\pi$, it is still possible to drop
the $\dl$ (smeared) piece of the OPEP while preserving gauge
invariance via a suitable modification
of the contact MEC current. Whether this should be done or not, however,
is a question (as is the question about which
value to choose for $\Lm_\pi$) which should
ultimately be answered by obtaining
the best possible agreement with experiment.
One might argue, however, that other mesons beyond the pion should be
included in the formalism before attempting such a task: we
intend to carry out this project in future work.

In closing this section, we wish to address the question of the
$k_F$--dependence  of the various pionic contributions to the nuclear
responses. In our view, $k_F$ is another parameter whose value should be set
by comparing the theoretical predictions with experimental data.
In figs.~17--20 we display the $k_F$--dependence of the various
contributions to the responses at $q=500$ MeV/c divided by the functions
$$K_L(\tau,\psi_r;\eta_F)=
{3{\cal N}\xi_F\over 4m_N\kappa\eta_F^3}{\kappa^2\over\tau}
\left[(1+\tau)W_2(\tau)-W_1(\tau)+W_2(\tau)\Delta\right]
\eqno(5.6a)$$
and
$$K_T(\tau,\psi_r;\eta_F)=
{3{\cal N}\xi_F\over 4m_N\kappa\eta_F^3}
\left[2W_2(\tau)+W_2(\tau)\Delta\right]
\eqno(5.6b)$$
for the longitudinal and transverse channels, respectively.
This permits us to focus only on the many--body part of the
responses, and these reduced quantities will eventually scale for
sufficiently high values of $q$.
{}From fig.~17, where we display the self--energy contribution
in the longitudinal electromagnetic channel, it is
apparent that as $k_F$ increases this contribution has a different
behaviour at high--$\omega$ than it does at
low--$\omega$ ({\it i.e.,\/} for scaling variable
$\psi_r >0$ and $<0$, respectively --- see ref.~\reff{16} for a discussion of
scaling in the RFG model). The net contribution to the total response,
however, remains quite small.

More pronounced is the sensitivity of the exchange terms to variations
of $k_F$, as
illustrated in figs.~18a and b where the $F$ and $B$ longitudinal
contributions are displayed.
Indeed one sees that the $F$--term grows with $k_F$ and
for $|\psi_r|\simeq 0.8$
yields, in the longitudinal channel, a contribution as large as about
25\% of the
free response when $k_F=250$~MeV/c (corresponding nuclei around $^{40}$Ca).
The $B$--term also grows with $k_F$,
although its impact on the nuclear response reaches only about
15\% of the free response.
As previously found, the transverse channel appears to be somewhat
less affected by the $F$--correlations (fig.~19). Also the $B$--correlations
are more
significant in the longitudinal channel than in the transverse one and,
overall, they are felt less than the $F$--correlations. The above results are
clearly in line with the dominance of the (smeared) $\delta$--term in shaping
the nuclear response as previously discussed.

As far as the MEC contributions (fig.~20) are concerned, their importance
increases with $k_F$ more rapidly in the
transverse channel  than in the longitudinal one. For small $k_F$
({\it i.e.,\/} either light nuclei or regions of low density)
they affect the nuclear responses only modestly; in particular, for
$k_F=$ 225 MeV/c they contribute only about 8\% and 3\% in
the transverse and longitudinal channels, respectively. At higher $k_F$
(250~MeV/c) they may yield a contribution larger than 10\%.

As a last point to be discussed in this section, in parallel to the issue
related to the impact of $\Lambda_\pi$ on the continuity equation discussed
above, we briefly
address the question of how global electromagnetic gauge invariance is affected
by the size of $k_F$. In this connection we consider the
longitudinal/charge\footnote{*}{\myrm The nomenclature used in studies of
electroweak  interactions with nuclei is not very good in this regard: what
is referred to as the charge response in some places and the longitudinal
response in others in fact usually involves both the $\mu=0$ (charge)
and $\mu=3$ or $z$ (longitudinal) projections of the 4--current. As we
discuss in this section, the two aspects are related by the continuity
equation when the current is conserved and thus it has become common
practice to use the terms interchangeably.  Throughout this article we have
used the word ``longitudinal'' to refer to all $\mu=0$ and 3 aspects of
electroweak current matrix elements.}
nuclear response, since it
is gauge invariance (through the continuity equation) that allows the
replacement of the longitudinal component of the spatial electromagnetic
current matrix elements with the charge
matrix elements, or {\it vice versa}.
We are not in a position to address this question at a quantitative level,
because in the present paper, for sake of simplicity, we have ignored the
convective current of the nucleon (only the dominant transverse spin current
has been kept). We can, however, offer a simple guiding principle to help in
recognizing whether gauge invariance is respected or not at the level of the
many--body longitudinal/charge response:
gauge invariance is fulfilled by a given
Goldstone diagram contributing to the polarization propagator if its two
electromagnetic vertices entail the {\it same} momentum flow.
To understand what this means it helps to look at fig.~21. From the examples
displayed there it is in fact apparent that, of the contributions calculated
in the present paper, the self--energy diagrams respect gauge invariance,
whereas the exchange ones do not. Analogously the
antisymmetrized RPA (and even the ring diagrams alone, Fig.~21b) and ladder
diagrams do not respect gauge invariance.

These differences may be understood by considering the general
expression for the response written as the sum of three terms, namely
$$ \left({Q^2\over q^2}\right)^2 R_L(q,\omega) =
R\left(QQ;q,\omega\right) - {2\omega\over q}R_z\left(QJ;q,\omega\right)
+{\omega^2\over q^2}R_{zz}\left(JJ;q,\omega\right) \ .
\eqno(5.7)$$
In the above $R\left(QQ;q,\omega\right)$, $R_z\left(QJ;q,\omega\right)$ and
$R_{zz}\left(JJ;q,\omega\right)$ refer to responses with two
charge, one charge and one longitudinal current and two longitudinal current
electromagnetic vertices, respectively.
The $z$--component of the nucleonic current, in the leading order of the
non--relativistic reduction, involves the convection current
$$J_z\propto(2\kb+\qb)\cdot\hat{\qb} \eqno(5.8)$$
and (2.1a), (3.1a) and (3.1b) contain integrals over $\kb$ (actually,
in the last two cases the integration is over a vector called $\kb_2$).
Therefore, by exploiting the energy--conserving $\delta$--function in the
self--energy contribution (2.1a), it turns out that the current (5.8) can be
expressed in this instance solely in terms of the external variables $q$ and
$\omega$. Clearly, the same does not occur in the exchange contributions
(3.1a) and (3.1b).
Thus, via the continuity equation, for a translationally--invariant system
treated at the level of the Hartree--Fock approximation, one has
$$ R_z\left(QJ;q,\omega\right) ={\omega\over q}R\left(QQ;q,\omega\right)
  \eqno(5.9a)$$
and
$$ R_{zz}\left(JJ;q,\omega\right) = {\omega^2\over q^2}
   R\left(QQ;q,\omega\right). \eqno(5.9b)$$
Hence
$$R_L(q,\omega) = R\left(QQ;q,\omega\right)\ , \eqno(5.10)$$
that is, $R_L(q,\omega)$ may be expressed entirely in terms of
a single response function.
This is not the case, for example, in the case of the antisymmetrized RPA.
It follows that for large $k_F$, where the exchange contribution becomes
relatively more
important, it might even be necessary to account for the whole
antisymmetrized RPA series, not to mention the ladder diagrams; in this
instance serious
consideration should be given to the gauge invariance of the theory.

\bigskip\bigskip\bigskip
\beginsection 6. Conclusions

Several motivations prompted the present study:
first, we wished to investigate the inclusive charge and spin electromagnetic
nuclear responses within a model that as much as possible respects Lorentz
and gauge invariance building on the work in ref.~\reff{15}.
Secondly, our intention has been to explore, in parallel with the above,
the weak neutral current responses (except for the axial--vector one whose
analysis
has been deferred to ref.~\reff{18}). We have thus paved the way to the
calculation of the asymmetry measured in parity--violating polarized
electron scattering, which shall also be dealt with in
ref.~\reff{18}.
Finally our purpose has been to investigate the degree to which a specific
nuclear model
based on nucleonic and mesonic degrees of freedom, when
treated consistently at the level of the forces and currents,
yields meaningful physical results over a relatively large range of
momentum transfers even when the framework assumed is
limited to first--order perturbation theory.

In this paper we have confined our attention to pionic effects, since the
scale governing the range over which the currents and interactions are
effective is especially large for the pion, being the lightest meson. Thus
we have started with pionic effects as our initial focus; this approach
serves as a paradigm for the
treatment of heavier mesons as well. Indeed, our basic approach in
which it has been possible to explore (albeit, within a specific
model of the nucleus) the interplay of forces and currents has
natural extensions where heavier mesons can be included as well.

Let us summarize our findings, beginning with the third of the motivations
mentioned above. We have strived
for consistency at two levels. The first one relates to the Feynman
diagrams: of these we have retained all cases having one pionic line.
This might appear insufficient for an adequate treatment of the forces and
yet we have been able to show that results obtained in first--order
perturbation theory are practically equivalent to those
obtained in
HF and RPA (especially to the first) when the force is carried by the pion.
Accordingly, we are led to conjecture
that even more sophisticated approaches, {\it e.g.,\/} RPA with
HF--dressed fermionic
propagators, would not be substantially different from those using
first--order perturbation
theory (at least in the 1p--1h sector).
This outcome is of importance because it would clearly be more difficult
to account for pionic MEC diagrams with two or more pionic lines.

The second level of consistency relates to the continuity equation.
Operationally this involves asking whether
the $\delta$--function part of the OPEP, given its crucial impact on the
nuclear responses,
affects the continuity equation or not. The answer is no for pointlike
nucleons and pions, and yes for extended ones.
Accordingly in the pointlike case one can, without affecting gauge invariance,
drop the $\delta$--function from the theory, arguing that short--range
correlations
will strongly quench its impact on the responses anyway.

On the other hand, some aspects of this quenching might be phenomenologically
embedded in
the factors $\Gamma_\pi$. But then the resulting {\it smeared} $\delta$
of the OPEP does contribute to the continuity equation. In such a case,
to fulfill the latter one needs to modify the currents beyond the standard
multiplicative factor $\Gamma_\pi^2$, indeed by adding
to the pion--in--flight current
two extra pieces (that vanish when the mass term $\Lambda_\pi\to\infty$).
Their contribution to the nuclear responses in the 1p--1h
sector is, however, quite small, although it should be checked whether this
remains true
in the 2p--2h sector and for heavier mesons.
A final option, not explored in the present paper, namely to drop
the smeared $\delta$ of the OPEP, might deserve further study.
In this case the restoration of gauge invariance can still be achieved by
 suitably  modifying the contact current, namely,
the only one contributed to by the $\delta$.

Turning to the second motivation, we have found a striking
correlation effect in the longitudinal weak neutral current response
$R^L_{AV}$ arising partly because the isovector and isoscalar components enter
in $R^L_{AV}$ with a particular sign combination (dictated by the standard
model), thus yielding a delicate cancellation in the case of the free RFG, and
partly because the $\tau=0$ pionic correlations are much stronger,
and of opposite sign, than the $\tau=1$ ones. This outcome has an important
bearing on the parity-violating polarized electron scattering asymmetry as
discussed in ref.~\reff{26} (see also ref.~\reff{18}).

Finally, coming to the first motivation above,
as far as the forces are concerned we have found that:
\smallskip
\item{i)} the charge and the spin responses are both hardened (shifted to
higher $\omega$) with respect
to the RFG;
\smallskip
\item{ii)} the hardening is greater in the longitudinal/charge channel (L)
than in the spin channel (T);
\smallskip
\item{iii)} the hardening fades away with increasing momentum transfer $q$
(at $\sim 1$~GeV/c very little is left of it);
\smallskip
\item{iiii)} the hardening is due to the $\Gamma_\pi$--modified
$\delta$--piece of the OPEP.

\bigskip
Regarding the currents (MEC) we have found that:
\smallskip
\item{i)} for the nuclear responses in the 1p--1h sector the MEC
contributions are small, but not negligible
(about 10\% of the nuclear response in the transverse channel and at most 5\%
in the longitudinal one);
\smallskip
\item{ii)} their contribution to the sum rules, relative to the RFG
contribution, is almost constant as a function of $q$;
\smallskip
\item{iii)} they always enter in such a way as to {\it decrease} the
nuclear responses.

In conclusion, our investigation of pionic effects started in ref.~\reff{15}
and pursued in the present paper shows the importance of the role they
play in determining the electromagnetic and weak neutral current
quasielastic nuclear responses. Predictions regarding the $q$--dependent
hardening of the longitudinal and transverse responses, $R_L$ and $R_T$,
as well as the dramatic modification of the longitudinal parity--violating
response $R_{AV}^L$ emerge from our results.  The approach taken, namely
through extensions to the relativistic Fermi gas model where Lorentz
covariance and consistency in treating forces and currents can be
maintained, allows us to explore the origin of these predictions in
a relatively direct way.  In particular, it has proven to be instructive
to see how they arise from the interplay of exchange and self--energy
contributions, of central and tensor pieces of the force and of
isoscalar/isovector correlation effects.  Naturally, more remains to be
done before a complete (hadronic) picture will be attained --- in future
work we intend to explore the nature of the quasielastic responses when
heavier mesons and summation schemes other than HF and RPA are
incorporated using the same basic RFG--motivated framework.

\bigskip\bigskip\bigskip

\beginsection Appendix  A

In this Appendix we perform the spin and isospin traces entering into
the self--energy and exchange diagrams of fig.~4.
\medskip
\noindent{\bf a) Spin}
\smallskip
The spin matrix element to be calculated in the first--order self--energy
term is
$$S_{L,T}^{s.e.}=\sum_{s_a s_b s_c s_d}
<s_a|\hat O^{\sigma}_{L,T}|s_b>
<s_b s_c|\hat V_{\sigma}|s_c s_d>
<s_d|\hat O^{\sigma}_{L,T}|s_a>\ ,
\eqno(A.1)$$
where $\hat O^{\sigma}_{L,T}$ is the spin operator associated to the
$\gamma$NN vertex in the longitudinal or transverse channel, namely
$$\hat O^{\sigma}_L = \onee
\eqno(A.2)$$
and
$$\hat O^{\sigma}_T = \left(\sgb\times\qbh\right)_i,
\eqno(A.3)$$
where $i=1,2$ and $\onee$ is the $2\times 2$ unit matrix.
In (A.1) $\hat V_{\sigma}$ can be either the central
($\sgb_1\cdot\sgb_2$) or the tensor operator
$${ S_{12}}(\kbh) =
(3\kh_i\kh_j - \dt_{ij})\ \sigma_{1i}\sigma_{2j}\ ,
\eqno(A.4)$$
$\kb$ being the momentum carried by the pion.
In the longitudinal channel one therefore gets
$$S_{L,central}^{s.e.}
=\sum_{s_a s_b s_c s_d} <s_a|\onee|s_b> <s_b|\sigma_i|s_c>
<s_c|\sigma_i|s_d> <s_d|\onee|s_a>  $$
$$=Tr\left\{\sigma_i\sigma_i\right\}=6,
\eqno(A.5)$$
(repeated indices are meant to be summed),
whereas the tensor force does not contribute to the
longitudinal self--energy:
$$S_{L,tensor}^{s.e.}=
\sum_{s_a s_b s_c s_d} <s_a|\onee|s_b> <s_b|\sigma_i|s_c>
<s_c|\sigma_j|s_d> <s_d|\onee|s_a> (3\kh_i\kh_j-\dt_{ij}) $$
$$=Tr\left\{\sigma_i\sigma_j\right\} (3\kh_i\kh_j-\dt_{ij}) = 0.
\eqno(A.6)$$

In the transverse channel the central part of the pion--exchange potential
yields
$$S_{T,central}^{s.e.}=
\sum_{s_a s_b s_c s_d} <s_a|(\sgb\times\qbh)_k|s_b> <s_b|\sigma_i|s_c>
<s_c|\sigma_i|s_d> <s_d|(\sgb\times\qbh)_k|s_a>  $$
$$=\varepsilon_{klm}\varepsilon_{krs}\qh_m\qh_s
Tr\left\{\sigma_l\sigma_i\sigma_i\sigma_r\right\} = 12.
\eqno(A.7)$$
Note that, in accord with the (A.3)
for $\hat O^{\sigma}_T$, although the sum over $k$ in (A.7) should
be restricted
to $k=1,2$, it is unnecessary to take all of the components, since
$$(\sgb\times\qbh)\cdot(\sgb\times\qbh) =
 (\sgb\times\qbh)_1 (\sgb\times\qbh)_1+
 (\sgb\times\qbh)_2 (\sgb\times\qbh)_2
 \eqno(A.8)$$
if $\qb$ points along the $z$--axis.  Using similar arguments one can
show that the
tensor interaction gives no contribution to the self--energy
in the transverse channel:

$$S_{T,tensor}^{s.e.}=$$
$$\sum_{s_a s_b s_c s_d} <s_a|(\sgb\times\qbh)_k|s_b> <s_b|\sigma_i|s_c>
<s_c|\sigma_j|s_d> <s_d|(\sgb\times\qbh)_k|s_a> (3\kh_i\kh_j-\dt_{ij})$$
$$=\varepsilon_{klm}\varepsilon_{krs}\qh_m\qh_s (3\kh_i\kh_j-\dt_{ij})
Tr\left\{\sigma_l\sigma_i\sigma_j\sigma_r\right\} =0,
\eqno(A.9)$$
having used the property
$$Tr\left\{\sigma_l\sigma_i\sigma_j\sigma_r\right\} =
2\left(\dt_{li}\dt_{jr}+\dt_{lr}\dt_{ij}-\dt_{lj}\dt_{ir}\right).
\eqno(A.10)$$

Let us now consider the exchange term. The spin factor is then given by:

$$S_{L,T}^{exch}=
\sum_{s_a s_b s_c s_d}
<s_a|\hat O^{\sigma}_{L,T}|s_b>
<s_b s_d|\hat V_{\sigma}|s_c s_a>
<s_c|\hat O^{\sigma}_{L,T}|s_d>.
\eqno(A.11)$$
In the longitudinal response this leads to
$$S_{L,central}^{exch}=
\sum_{s_a s_b s_c s_d} <s_a|\onee|s_b> <s_b|\sigma_i|s_c>
<s_d|\sigma_i|s_a> <s_c|\onee|s_d>  $$
$$=Tr\left\{\sigma_i\sigma_i\right\}=6
  \eqno(A.12)$$
and
$$S_{L,tensor}^{exch}
  =\sum_{s_a s_b s_c s_d} <s_a|\onee|s_b> <s_b|\sigma_i|s_c>
<s_d|\sigma_j|s_a> <s_c|\onee|s_d> (3\kh_i\kh_j-\dt_{ij}) $$
$$=Tr\left\{\sigma_i\sigma_j\right\} (3\kh_i\kh_j-\dt_{ij}) = 0.
\eqno(A.13)$$
In the transverse response
$$S_{T,central}^{exch}
  =\sum_{s_a s_b s_c s_d} <s_a|(\sgb\times\qbh)_k|s_b> <s_b|\sigma_i|s_c>
<s_d|\sigma_i|s_a> <s_c|(\sgb\times\qbh)_k|s_d>$$
$$=\varepsilon_{klm}\varepsilon_{krs}\qh_m\qh_s
Tr\left\{\sigma_l\sigma_i\sigma_r\sigma_i\right\} = -4
\eqno(A.14)$$
and
$$S_{T,tensor}^{exch}=$$
$$\sum_{s_a s_b s_c s_d} <s_a|(\sgb\times\qbh)_k|s_b>
<s_b|\sigma_i|s_c>
<s_d|\sigma_j|s_a> <s_c|(\sgb\times\qbh)_k|s_d>
(3\kh_i\kh_j-\dt_{ij})$$
$$=\varepsilon_{klm}\varepsilon_{krs}\qh_m\qh_s (3\kh_i\kh_j-\dt_{ij})
Tr\left\{\sigma_l\sigma_i\sigma_r\sigma_j\right\} =
4\left[1-3\left(\kbh\cdot\qbh\right)^2\right].
\eqno(A.15)$$
Note that the transverse channel is the only one where the tensor interaction
gives a nonzero contribution, namely, via the exchange diagram.
\medskip
\noindent{\bf b) Isospin}
\smallskip
Let us now consider the isospin matrix elements involved in the self--energy
and exchange terms.
At the $\gamma$NN vertex we have both isoscalar and isovector
($z$--component) dependences: for example, the charge and convection current
operators involve
$$\hat O = {\onee+\tau_z \over 2}
\eqno(A.16a)$$
and the magnetization operator involves
$$\eqalign{\hat O &= \Bigl[{\onee+\tau_z \over 2}\Bigr]\mu_p +
\Bigl[{\onee-\tau_z \over 2}\Bigr]\mu_n \cr
&= \mus\ {\onee\over 2} + \muv\ {\tau_z\over 2}\ . \cr}\eqno(A.16b)$$
Here the $\onee$--terms give rise to the isoscalar ($\tau=0$)
particle--hole responses, while the $\tau_z$--terms correspond to the
isovector ($\tau=1$) ones.

The self--energy diagram takes equal contributions from the traces in
the two isospin channels:
$$T^{s.e.}(\tau=0)={1\over 4}\sum_{t_a t_b t_c t_d}
<t_a|\onee|t_b> <t_b t_c|{\bf\tau_1}\cdot{\bf\tau_2}|t_c t_d>
<t_d|\onee|t_a>$$
$$={1\over 4}Tr\{\tau_i\tau_i\} = {3\over 2}
\eqno(A.17)$$
and
$$T^{s.e.}(\tau=1)={1\over 4}\sum_{t_a t_b t_c t_d}
<t_a|\tau_z|t_b> <t_b t_c|{\bf\tau}_1\cdot{\bf\tau}_2|t_c t_d>
<t_d|\tau_z|t_a>$$
$$={1\over 4}Tr\{\tau_z\tau_i\tau_i\tau_z\} = {3\over 2}.
  \eqno(A.18)$$
In contrast, in the exchange term the isoscalar and the isovector
traces are different in magnitude and sign. They read
$$T^{exch}(\tau=0)={1\over 4}\sum_{t_a t_b t_c t_d}
<t_a|\onee|t_b> <t_b t_d|{\bf\tau}_1\cdot{\bf\tau}_2|t_c t_a>
<t_c|\onee|t_d>$$
$$  ={1\over 4}Tr\{\tau_i\tau_i\} = {3\over 2}
\eqno(A.19)$$
and
$$T^{exch}(\tau=1)={1\over 4}\sum_{t_a t_b t_c t_d}
<t_a|\tau_z|t_b> <t_b t_d|{\bf\tau}_1\cdot{\bf\tau}_2|t_c t_a>
<t_c|\tau_z|t_d>$$
$$  ={1\over 4}Tr\{\tau_z\tau_i\tau_z\tau_i\} = -{1\over 2}.
 \eqno(A.20)$$

Finally let us consider the $n^{\rm th}$--order exchange diagram, which
contains $n$ pionic lines inside one bubble as in fig.~7.
The corresponding isospin factor reads
$$T_n^{exch}=
Tr\{\hat O^{\tau}\tau_{i1}\tau_{i2}...\tau_{in-1}
\tau_{in} \hat O^{\tau}\tau_{in}\tau_{in-1}...\tau_{i2}\tau_{i1}\}
\eqno(A.21)$$
and can be obtained from the first--order ones by recursion relations:
$$T_n^{exch}(\tau=0)= {\ 3^n\over 2}
\eqno(A.22)$$
and
$$T_n^{exch}(\tau=1)= -T_{n-1}^{exch}(\tau=1) = {(-1)^n\over 2}.
\eqno(A.23)$$

\bigskip\bigskip\bigskip

\beginsection Appendix  B

In this Appendix we first give the functions $\Nc_{L,T}$ and $\Hc_T$ entering
in the formulae (3.1-4) for the exchange contributions to the response.
The first two functions are the following:
$$\eqalignno{
  &\Nc_L(k_1,k_2;z,\psi) = {\lmt^2\over\mpt^2}(\lmt^2-\mpt^2)
     {4k_1k_2\over \Fc^2_{\lmt}(k_1,k_2;\psi,\psi)}\cr
  &\times\left\{{\lmt^2+k_1^2+k_2^2-2\psi^2\over
     \Fc^2_{\lmt}(k_1,k_2;\psi,\psi)}
     \left[{\rm ln}\big|\Gc_{\lmt}(k_1,k_2;z,\psi)\big|-
     {\rm ln}|\psi-z|\right]\right.+1\cr
  &\left.-
   { (\lmt^2+k_1^2+k_2^2)^3-4k_1^2k_2^2(\lmt^2+k_1^2+k_2^2)
    -4\psi^2k_2^2(\lmt^2-k_1^2+k_2^2)-4\psi z k_1^2(\lmt^2+k_1^2-k_2^2)\over
   [(\lmt^2+k_1^2+k_2^2)^2-4k_1^2k_2^2]\Fc_{\lmt}(k_1,k_2;z,\psi)}\right\}\cr
  &\qquad+\Mc_L^{\mpt}(k_1,k_2;z,\psi)-\Mc_L^{\lmt}(k_1,k_2;z,\psi)
  &(B.1)\cr}$$
and
$$\eqalignno{
  &\Nc_T(k_1,k_2;z,\psi) = -{(\lmt^2-\mpt^2)\over\mpt^2}
     {k_2\over k_1}\left\{{\psi\over k_1}{\rm ln}\big|2k_1^2z
     -\psi(\lmt^2+k_1^2+k_2^2)+k_1\Fc_{\lmt}(k_1,k_2;z,\psi)\big|\right.\cr
  &\quad\left.+{(\lmt^2+k_1^2+k_2^2)^3
     -4(\lmt^2+k_1^2+k_2^2)(k_1^2k_2^2+\lmt^2\psi z)
     +4\psi k_2^2(\psi-z)(\lmt^2-k_1^2+k_2^2)\over
     [(\lmt^2+k_1^2+k_2^2)^2-4k_1^2k_2^2]\Fc_{\lmt}(k_1,k_2;z,\psi)}\right\}
  \cr
  &\quad+\Mc_T^{\mpt}(k_1,k_2;z,\psi)-\Mc_T^{\lmt}(k_1,k_2;z,\psi),
  &(B.2)\cr}$$
where
$$\eqalign{
  \Mc_L^\lm(k_1,k_2;z,\psi) &= -{4k_1k_2\over\Fc_\lm(k_1,k_2;\psi,\psi)}
   \left[{\rm ln}\big|\Gc_\lm(k_1,k_2;z,\psi)\big|-{\rm ln}|\psi-z|\right]\cr
  \Mc_T^\lm(k_1,k_2;z,\psi) &= -{1\over\mpt^2}{k_2\over k_1}
     \big\{\Fc_\lm(k_1,k_2;z,\psi)\cr
   + {\psi\over k_1}&(\lm^2-k_1^2+k_2^2){\rm ln}\big|2k_1^2z
     -\psi(\lm^2+k_1^2+k_2^2)+k_1\Fc_{\lm}(k_1,k_2;z,\psi)\big|\big\} \cr
  \Gc_\lm(k_1,k_2;z,\psi) &= 2\psi(\psi-z)(\lm^2-k_1^2+k_2^2)
     +\Fc_\lm(k_1,k_2;z,\psi) \Fc_\lm(k_1,k_2;\psi,\psi)\cr
     &\qquad+\Fc_\lm^2(k_1,k_2;\psi,\psi) \cr
  \Fc_\lm(k_1,k_2;z,\psi) &= \sqrt{(\lm^2+k_1^2+k_2^2)^2
     -4\psi z(\lm^2+k_1^2+k_2^2)+4k_1^2z^2+4k_2^2\psi^2-4k_1^2k_2^2}.\cr}
  \eqno(B.3)$$
We have also set $\lmt=\Lambda_\pi/k_F$ and $\mpt=m_\pi/k_F$.
The function $\Hc_T$, which gives in a fully analytic form the
transverse exchange response at $q>2k_F$, reads
$$\eqalignno{
  \Hc_T(k,p) &= -{(\lmt^2-\mpt^2)\over\mpt^2}\left\{2\lmt\left[
     \tan^{-1}\left({k+p-1\over\lmt}\right)
     -\tan^{-1}\left({k-p+1\over\lmt}\right)\right]\right.\cr
     &\qquad\left.+{1\over2k}[\lmt^2-k^2+(p-1)^2]
     {\rm ln}\left|{\lmt^2+(k-p+1)^2\over\lmt^2+(k+p-1)^2}\right|\right\}\cr
   &\qquad+\Tc^{\mpt}(k,p)-\Tc^{\lmt}(k,p), &(B.4)\cr}$$
where
$$\eqalignno{
  \Tc^\lm(k,p) &= {1\over\mpt^2}\left\{{k^2\over3}(p-1)
     +{4\over3}\lm^3\left[ \tan^{-1}\left({k-p+1\over\lm}\right)
     -\tan^{-1}\left({k+p-1\over\lm}\right)\right]\right.\cr
     &\qquad\quad
     +{1\over12k}[k^4-6k^2(1-\lm^2)-3(1+\lm^2)^2+12p(\lm^2+k^2+1)\cr
     &\qquad\qquad\quad-6p^2(\lm^2+k^2+3)-3p^3(p-4)]
     {\rm ln}\left|{\lmt^2+(k-p+1)^2\over\lmt^2+(k+p-1)^2}\right|\cr
     &\qquad\quad\left.
     +{2\over3}(p-1)^3
     {\rm ln}\left|[\lmt^2+(k-p+1)^2][\lmt^2+(k+p-1)^2]\right|\right\}.
     &(B.5)\cr}$$
\medskip
Finally, we give the expression for the real part of the first--order
forward--going ($F$)
longitudinal polarization propagator (see eq.~(3.6)), limited for simplicity
to the case $q>2 k_F$:
$$\eqalignno{
  &{\rm Re}\Pi^{(1){\rm F}}_{L,\tau}(q,\om) = \xi_A{1\over k_F^3}
    {f^2_\pi\over m_\pi^2} {\cal I}_\tau
  \int\! {d\kb_1\over(2\pi)^3}{d\kb_2\over(2\pi)^3}
    \theta(k_F-k_1) \theta(k_F-k_2) {(\kb_1-\kb_2)^2
    \over m_\pi^2+(\kb_1-\kb_2)^2}   \cr
  &\qquad\times\Bigg[
    {1\over\om-{|Q^2|/2m_N}-{\qb\cdot\kb_1/ m_N}}\;
    {1\over\om-{|Q^2|/2m_N}-{\qb\cdot\kb_2/ m_N}} \cr
  &\qquad\quad -
    \dl(\om-{|Q^2|/2m_N}-{\qb\cdot\kb_1/ m_N})\,
    \dl(\om-{|Q^2|/2m_N}-{\qb\cdot\kb_2/ m_N}) \Bigg]
      &(B.6a) \cr
  &\phantom{{\rm Re}\Pi^{(1){\rm TD}}_{L,\tau}(q,\om)}
  ={1\over128\pi^4}{\xi_A\over\eta_F\kappa^2}
  {f_\pi^2\over m_N}{\cal I}_\tau\cr
  &\qquad\qquad\times\Bigg\{\int_0^1\!dk_1\int_{-1}^1\!dx_1\int_0^1\!dk_2
  \big[{\cal P}(k_1,k_2;k_1x_1,\psi_r)
  -{\cal P}(k_1,-k_2;k_1x_1,\psi_r)\big]\cr
  &\qquad\qquad\quad +\pi^2\int_{|\psi_r|}^1\!dk\,k{\rm ln}\left|
    {1+\mpt^2-k^2+\sqrt{(1-\mpt^2-k^2)^2+4\mpt^2(1-\psi_r^2)} \over
    1+\lmt^2-k^2+\sqrt{(1-\lmt^2-k^2)^2+4\lmt^2(1-\psi_r^2)}}\right|\cr
  &\qquad\qquad\quad
    - {\lmt^2-\mpt^2\over \mpt^2}{\pi^2\over4}\big(\lmt^2+2
    -2\psi_r^2-\lmt\sqrt{\lmt^2+4-4\psi_r^2}\big)\cr
  &\qquad\qquad\quad
    -{\pi^2\over2}(1-\psi_r^2){\rm ln}\left({\mpt^2\over\lmt^2}\right)\Bigg\},
  &(B.6b)\cr}$$
where ${\cal I}_{\tau=0}=3/2$, ${\cal I}_{\tau=1}=-1/2$,
$$\eqalignno{
  {\cal P}(k_1,k_2;y_1,\psi) &= {\lmt^2\over \mpt^2}(\lmt^2-\mpt^2)
    \left\{{-1\over\psi-y_1}{4k_1^2k_2^2[y_1(\lmt^2+k_1^2+k_2^2)-2k_1^2\psi]
    \over[(\lmt^2+k_1^2+k_2^2)^2-4k_1^2k_2^2]\Fc_{\lmt}^2(k_1,k_2;y_1,\psi)}
    \right.\cr
  &\qquad\qquad\left.
  -{\lmt^2+k_1^2+k_2^2-2y_1\psi\over\Fc_{\lmt}^2(k_1,k_2;y_1,
  \psi)}{\cal Q}^{\lmt}(k_1,k_2;y_1,\psi)\right\}\cr
  &\quad + {\cal Q}^{\mpt}(k_1,k_2;y_1,\psi)
  -{\cal Q}^{\lmt}(k_1,k_2;y_1,\psi)
    &(B.7a)\cr}$$
and
$$\eqalignno{
  {\cal Q}^\lm(k_1,k_2&;y_1,\psi) = -{k_1^2k_2\over\psi-y_1}
    {1\over\Fc_\lm(k_1,k_2;y_1,\psi)} \cr
  &\quad\times\big\{ {\rm ln}|\psi+k_2|+
   {\rm ln}\big|2y_1(\psi-k_2)
   (\lm^2-k_1^2+k_2^2)-4k_1^2(\psi-k_2)(\psi-y_1)\cr
  &\qquad\qquad + (\lm^2+k_1^2+k_2^2-2y_1k_2)\Fc_\lm(k_1,k_2;y_1,\psi)+
    \Fc_\lm^2(k_1,k_2;y_1,\psi)\big| \big\}.\cr&&(B.7b)\cr}$$

\bigskip\bigskip\bigskip

\beginsection References

\smallskip
\medskip
\item{[1]}  T. Ericson and W. Weise, {\it Pions and Nuclei}
(Clarendon Press, Oxford, 1988)
\smallskip
\medskip
\item{[2]} R.A.M. Klomp, V.G.J. Stoks and J.J. de~Swartt, Phys. Rev.
{\bf C44} (1991) 1258
\smallskip
\medskip
\item{[3]} T.E.O. Ericson and M. Rosa--Clot, Ann. Rev. Nucl. Part. Sci.
{\bf 35} (1985) 271
\smallskip
\medskip
\item{[4]} D.O. Riska and G.E. Brown, Phys. Lett. {\bf B38} (1972) 193
\smallskip
\medskip
\item{[5]} J. Hockert {\it et al.}, Nucl. Phys. {\bf A217} (1973) 19
\smallskip
\medskip
\item{[6]}  W. Fabian and H. Arenh\"ovel, Nucl. Phys. {\bf A314} (1979) 253
\smallskip
\medskip
\item{[7]}  J.L. Friar, B.F. Gibson and G.L. Payne, Phys. Rev.
{\bf C30} (1984) 1084; J.L. Friar, in {\it Proceedings of the Fifth
Indian--Summer School on Intermediate Energy Physics,} S\'azava,
Czechoslovakia, 1992
\smallskip
\medskip
\item{[8]} R.B. Wiringa, Phys. Rev. {\bf C43} (1991) 1585
\smallskip
\medskip
\item{[9]} W.T. Weng, T.T.S. Kuo and G.E. Brown, Phys. Lett. {\bf B46}
(1973) 329
\smallskip
\medskip
\item{[10]} E. Oset, H. Toki and W. Weise, Phys. Rep. {\bf 83} (1982) 281
\smallskip
\medskip
\item{[11]} W.M. Alberico, M. Ericson and A. Molinari, Phys. Lett.
{\bf 92B} (1980) 153
\smallskip
\medskip
\item{[12]} J. Dubach, J.H. Koch and T.W. Donnelly, Nucl. Phys. {\bf A271}
(1976) 279
\smallskip
\medskip
\item{[13]} B.D. Serot and J.D. Walecka, Adv. in Nucl. Phys. {\bf 16},
eds. J.W. Negele and E. Vogt (Plenum Press, New York, 1986)
\smallskip
\medskip
\item{[14]} V.R. Pandharipande, in {\it Proceedings of the International
Nuclear Physics Conference,\/} Wiesbaden, Germany (1992)
\smallskip
\medskip
\item{[15]} W.M. Alberico, T.W. Donnelly and A. Molinari, Nucl. Phys.
{\bf A512} (1990) 541
\smallskip
\medskip
\item{[16]} W.M. Alberico {\it et al.}, Phys. Rev. {\bf C38} (1988) 1801
\smallskip
\medskip
\item{[17]} J.F. Mathiot, Phys. Rep. {\bf 173} (1989) 63
\smallskip
\medskip
\item{[18]} W.M. Alberico {\it et al.}, in preparation
\smallskip
\medskip
\item{[19]} W.M. Alberico, R. Cenni and A. Molinari, La Rivista del Nuovo
Cimento {\bf 1} (1978) 1
\smallskip
\medskip
\item{[20]}  R. Rosenfelder, Phys. Lett. {\bf 79B} (1978) 15
\smallskip
\medskip
\item{[21]} P. Ring and P. Schuck, {\it The Nuclear Many--Body Problem}
(Springer--Verlag, Berlin, 1980)
\smallskip
\medskip
\item{[22]} A. Dellafiore, F. Lenz and F.A. Brieva, Phys. Rev. {\bf C31}
(1985) 1088; F. Lenz, E.J. Moniz and K. Yazaki, Ann. Phys. {\bf 129}
(1980) 84
\smallskip
\medskip
\item{[23]}  T.W. Donnelly and R.D. Peccei, Phys. Rep. {\bf 50} (1979) 1
\smallskip
\medskip
\item{[24]}  M.J. Musolf and T.W. Donnelly, Nucl. Phys. {\bf A546} (1992) 509
\smallskip
\medskip
\item{[25]}  M.J. Musolf and T.W. Donnelly, to be published in Z. f\"ur Physik
\smallskip
\medskip
\item{[26]}  T.W. Donnelly {\it et al.}, Nucl. Phys. {\bf A541} (1992) 525
\smallskip
\medskip
\item{[27]}  E. Hadjimichael, G.I. Poulis and T.W. Donnelly, Phys. Rev.
{\bf C45} (1992) 2666

\bigskip\bigskip\bigskip

\beginsection Figure Captions

\medskip
\item{}
Fig.~1. Goldstone diagrams for the self--energy, with the pion dressing
the particle (a) or the hole (b) line.
\medskip
\item{}
Fig.~2. The self--energy contribution to the longitudinal response function
is shown as a function of the energy transfer $\omega$ for
three values of the momentum transfer: $q$=300, 500, and 1000 MeV/c.
The dashed lines represent the separated particle and hole terms,
the solid line being their sum.
Here and in the following figures, unless otherwise specified,
the nucleus considered is $^{12}$C,
$k_F$=225 MeV/c and $\Lambda_\pi$=1.3 GeV.
\medskip
\item{}
Fig.~3. The Hartree--Fock longitudinal response (solid curve) is displayed
and compared to the free RFG (dotted) and first--order self--energy
correlated (dashed) responses.  The first and last almost coincide.
\medskip
\item{}
Fig.~4. Goldstone diagrams corresponding to the forward--going (a) and
backward--going (b) terms of the exchange correlations, defined in
eqs.~(3.3) and (3.4).  In the text these are labeled ``$F$'' and ``$B$'',
respectively.
\medskip
\item{}
Fig.~5. The exchange contribution to the longitudinal response is shown
(solid curves). The forward-- (dashed) and backward--going (dot--dashed)
terms, ``$F$'' and ``$B$'', respectively, are also separately displayed.
\medskip
\item{}
Fig.~6. The same as fig.~5, but now for the transverse response.
\medskip
\item{}
Fig.~7. Goldstone diagrams entering into the Tamm--Dancoff series for
the polarization propagator with pion exchange.
\medskip
\item{}
Fig.~8. The free (dotted), first--order exchange (dashed) and first
continued fraction (solid) longitudinal responses are plotted as
functions of $\omega$ at $q$=500 and 1000 MeV/c.
For the results in the left--hand panel, the interaction is the pure
``$\delta$--like'' part of the pion potential, while for the results in the
right--hand panel the full potential is taken.
\medskip
\item{}
Fig.~9. The free Fermi gas longitudinal response is shown as dotted curves;
the dashed curves include the self--energy and exchange correlations,
whereas the solid curves also include MEC.
\medskip
\item{}
Fig.~10. The same as fig.~9, but now for the transverse response.
\medskip
\item{}
Fig.~11. The shift of the quasielastic peak due to pionic correlations
is shown as a function of $q$ in the longitudinal (solid curve) and transverse
(dashed curve) channels.
\medskip
\item{}
Fig.~12. The longitudinal (upper panel) and transverse (lower panel)
sum rules $S_L(q)$ and $S_T(q)$, as defined through
eq.~(4.5), are shown as functions of $q$. The dotted curves correspond
to the RFG, the dashed curves include self--energy and exchange correlations
and the solid ones contain as well the MEC contribution.
In the small windows the ratios of the correlated sum rules to the free
ones are displayed: the dashed curves do not include the MEC, the solid
ones do.
\medskip
\item{}
Fig.~13. Upper panel: the isoscalar free (dashed) and correlated (solid)
longitudinal responses are displayed as functions of $\omega$ at $q$=500 MeV/c.
Lower panel: the same quantities are shown in the isovector channel.
\medskip
\item{}
Fig.~14. The longitudinal weak neutral current response $R^L_{AV}$ at $q$=
500 MeV/c, showing the free RFG (dashed) and correlated (solid) results.
\medskip
\item{}
Fig.~15. The exchange contribution to the longitudinal response at
$q$=500 MeV/c is displayed in the upper panel for three different values of
$\Lambda_\pi$:
10 GeV (dashed), 1.3 GeV (dot--dashed) and 0.8 GeV (dotted). In the lower
panel the longitudinal self--energy is shown, with the same meaning for the
three curves.
\medskip
\item{}
Fig.~16. The exchange (upper panel) and MEC (lower panel) contributions to
the transverse
response at $q$=500 MeV/c are displayed for $\Lambda_\pi$=
10 GeV (dashed), 1.3 GeV (dot--dashed) and 0.8 GeV (dotted).
\medskip
\item{}
Fig.~17. The self--energy contribution to the longitudinal response
divided by the function $K_L(\tau,\psi_r;\eta_F)$ (eq.~5.6a)
at $q$=500 MeV/c is displayed as a function of the scaling variable
$\psi_r$; three different values of $k_F$ have been chosen:
200 MeV/c (dashed), 225 MeV/c (dot--dashed) and 250 MeV/c (solid).
\medskip
\item{}
Fig.~18. The forward--going ($F$, upper panel) and backward--going
($B$, lower panel) terms of the
exchange, divided by $K_L(\tau,\psi_r;\eta_F)$ in the longitudinal channel
shown as functions of $\psi_r$ for $k_F$=200 MeV/c (dashed), 225 MeV/c
(dot--dashed) and 250 MeV/c (solid) at $q$=500 MeV/c.
\medskip
\item{}
Fig.~19. The forward--going ($F$, upper panel) and backward--going
($B$, lower panel) terms of the
exchange, divided by $K_T(\tau,\psi_r;\eta_F)$ (eq.~5.6b)
in the transverse channel are shown as functions of $\psi_r$
for $k_F$=200 MeV/c (dashed), 225 MeV/c (dot--dashed) and 250 MeV/c (solid)
at $q$=500 MeV/c.
\medskip
\item{}
Fig.~20. The MEC contribution to the longitudinal response divided
by $K_L(\tau,\psi_r;\eta_F)$ is displayed in the upper panel as a function of
$\psi_r$
for $k_F$=200 MeV/c (dashed), 225 MeV/c (dot--dashed) and 250 MeV/c (solid)
at $q$=500 MeV/c. In the lower panel the transverse MEC contribution divided
by $K_T(\tau,\psi_r;\eta_F)$ is shown for the same set of $k_F$ values.
\medskip
\item{}
Fig.~21. Gauge invariance in many--body theories.
Diagram (a) (self--energy) has the same momentum flow in the two
electromagnetic vertices.
Diagrams (b) (ring), (c) (exchange), (d) (exchange contributions to
antisymmetrized RPA) and (e)
(ladder) have not. The first class of diagrams fulfills gauge invariance, the
second does not.
\vfill
\bye